\documentclass[12pt,a4paper]{article}
\usepackage{cite}
\pdfoutput=1
\usepackage{graphicx}
\usepackage[T1]{fontenc}
\usepackage[utf8]{inputenc}
\usepackage{textcomp}
\usepackage[sc,osf]{mathpazo}
\usepackage{a4wide}
\usepackage{multirow}
\usepackage{multicol}
\usepackage{latexsym,amsthm,amsfonts,amsmath,mathrsfs,amssymb}

\usepackage{accents}

\usepackage{booktabs} 
\usepackage[unicode,implicit]{hyperref}
\newcommand{\EB}[1]{\textcolor{red}{#1}}
\hypersetup{%
  pdftitle    = {Lie Algebra Expansions and Actions for Non-Relativistic Gravity}
  pdfkeywords = {gravity, Newton-Cartan, Galileo, Wigner-Inonu contractions, Lie algebra expansions, Einstein-Hilbert action, first-order action}
  pdfauthor   = {Eric Bergshoeff, Jose M. Izquierdo, Tomas Ortin, Luca Romano},
  plainpages  = true,
  colorlinks  = true,
  citecolor   = blue,
  urlcolor    = red,
  linkcolor   = black
}
\newcommand{\hepth}[1]{{\tt
\href{http://www.arXiv.org/abs/hep-th/#1}{hep-th/#1}}}

\newcommand{\arxiv}[1]{{\tt arXiv:\href{http://www.arXiv.org/abs/#1}{#1}}}
\newcommand{\doi}[1]{{\tt DOI:\href{http://dx.doi.org/#1}{#1}}}

\makeatletter
\@addtoreset{equation}{section}
\makeatother

\begin{document}

\begin{flushright}
\small
IFT-UAM/CSIC-19-10\\
\normalsize
\end{flushright}

\begin{center}

  {\Large {\bf Lie Algebra Expansions and \\[.5cm]
      Actions for Non-Relativistic Gravity}}

\vspace{1.5cm}

\renewcommand{\thefootnote}{\alph{footnote}}
{\sl\large Eric Bergshoeff$^{~1}$}\footnote{Email: {\tt e.a.bergshoeff[at]rug.nl}},
{\sl\large Jos\'e Manuel Izquierdo$^{~2}$}\footnote{Email: {\tt izquierd[at]fta.uva.es}},
{\sl\large Tom\'{a}s Ort\'{\i}n$^{~3}$}\footnote{Email: {\tt tomas.ortin[at]csic.es}}\\[.5cm]
{\sl\large and Luca Romano$^{~3}$}\footnote{Email: {\tt lucaromano2607[at]gmail.com}}

\setcounter{footnote}{0}
\renewcommand{\thefootnote}{\arabic{footnote}}

\vspace{0.5cm}

${}^{1}${\it Van Swinderen Institute, University of Groningen,\\ Nijenborgh 4,
  9747 AG Groningen, The Netherlands}

\vspace{0.2cm}

${}^{2}${\it  Departamento de F\'{\i}sica Te\'orica, Universidad de Valladolid,\\
  E-47011-Valladolid, Spain}

\vspace{0.2cm}

${}^{3}${\it Instituto de F\'{\i}sica Te\'orica UAM/CSIC\\
C/ Nicol\'as Cabrera, 13--15,  C.U.~Cantoblanco, E-28049 Madrid, Spain}\\

\vspace{.8cm}


{\bf Abstract}
\end{center}
\begin{quotation}
  {\small We show that the general method of Lie algebra expansions
    can be applied to re-construct several algebras and related
    actions for non-relativistic gravity that have occurred in the
    recent literature. We explain the method and illustrate its
    applications by giving several explicit examples. The method can
    be generalized to include the construction of actions for
    ultra-relativistic gravity, i.e.~Carroll gravity, and
    non-relativistic supergravity as well.  }
\end{quotation}

\newpage
\pagestyle{plain}

\tableofcontents


\section{Introduction}

During the last few years, there has been a renewed interest in
non-relativistic gravity due to its possible role in holography where
gravity is used as a tool to probe the dark corners of quantum field
theories\footnote{For a discussion of non-relativistic gravity in the
  bulk, see \cite{Bagchi:2009my}; for an application of
  non-relativistic gravity at the boundary, see
  \cite{Christensen:2013lma}.}  or in effective field theory where
gravitational fields are used as geometrical background response
functions \cite{Son:2013rqa,Geracie:2014nka}.

It turns out that, in contrast to general relativity, there are many
different versions of non-relativistic gravity theories.  They are all
invariant under reparametrizations but differ in the fact that they
are invariant under distinct extensions of the Galilei symmetries. The
simplest example is the so-called Galilei gravity theory which is
invariant under the un-extended Galilei symmetries
\cite{Bergshoeff:2017btm}.  Instead, Newtonian gravity and its
frame-independent reformulation, called Newton-Cartan (NC) gravity, is
invariant under the symmetries corresponding to a central extension of
the Galilei algebra, called the Bargmann algebra.  In three dimensions
there even exists a non-relativistic gravity theory that is based on a
Galilei algebra with {\sl two} central extensions
\cite{Papageorgiou:2009zc,Bergshoeff:2016lwr,Hartong:2016yrf} while
recently, in four dimensions, an interesting example of a
non-relativistic gravity theory based upon an even larger extension of
the Bargmann algebra containing non-central generators has been
encountered \cite{Hansen:2018ofj}.

It is well known that the generator of time translations of the
Galilei algebra plays a different role than the generators of space
translations. This is related to the fact that absolute time plays a
special role in Newtonian gravity. However, it turns out that this
type of Galilei algebra is only suitable when coupling
non-relativistic gravity to particles. Instead, when coupling to
extended objects with $p$ spatial extensions,
\textit{i.e.}~$p$-branes, one needs to foliate spacetime in a
different way, according to the number of \textit{longitudinal} and
\textit{transverse} directions (directions parallel to the $p$-brane's
worldvolume and the rest). This leads to a different kind of Galilei
algebra where, for a given spacetime of dimension $D$, the translation
generators of the Galilei algebra are given by the following set of
generators
\begin{equation}
(H_{A}, P_{a})\,, \hskip 1truecm A=0,1,\cdots, p;\qquad  a=p+1,\cdots, D-1\,,
\end{equation}
which correspond to translations along longitudinal and transverse
directions, respectively. We will denote these different kinds of
Galilei algebra as {\sl $(D,p)$ Galilei algebras}. The related
non-relativistic gravity theories will be called {\sl $(D,p)$ Galilei
  gravity}.\footnote{From now on it is understood that, whenever the value of $p$ is not specified, we  mean the particle case,
  \textit{i.e.}~$p=0$.}  Like in the particle case, there are further
non-relativistic gravity theories that are based upon {\sl extensions}
of the $(D,p)$ Galilei algebras.  A recent example is string NC
gravity \cite{Andringa:2012uz} that is based upon a $p=1$ (string)
version of the Bargmann algebra \cite{Brugues:2004an}.\footnote{The
  action for a non-relativistic Polyakov string in such a string NC
  background has been recently constructed \cite{Bergshoeff:2018yvt}
  thereby generalizing the action of \cite{Gomis:2000bd}.}

It is a relatively simple matter to construct an action for $(D,p)$
Galilei gravity.  One simply takes an appropriate non-relativistic
limit of the first-order Einstein-Hilbert action, thereby generalizing
the case of Galilei gravity discussed in
\cite{Bergshoeff:2017btm}.\footnote{In this work we will only consider
  first-order actions with independent spin connection fields. We will
  comment about second-order formulations in the conclusions.}  It is
much harder to construct an action for NC gravity that is based on the
Bargmann algebra. Taking the non-relativistic limit of general
relativity without obtaining fatal divergencies can only be done at
the level of the equations of motion. Similarly, performing a
null-reduction of five-dimensional general relativity does not lead to
an action for 4-dimensional NC gravity but only to its equations of
motion \cite{Julia:1994bs}.  Despite these difficulties, several
actions for non-relativistic gravity, based on extensions of the
Galilei algebra, have been constructed in the recent literature. One
example we already mentioned is the three-dimensional action of
\cite{Papageorgiou:2009zc,Bergshoeff:2016lwr,Hartong:2016yrf}, which
is based upon a three-dimensional Galilei algebra with two central
central extensions.  More recent examples are the 4D non-relativistic
gravity action of \cite{Hansen:2018ofj}, the 4D extended string
Newton-Cartan gravity action of \cite{Bergshoeff:2018vfn} and the 3D action of \cite{Ozdemir:2019orp}.\,\footnote{In this paper the word `extended'  will be used to denote certain extended (super)-gravity theories, as in the mathematical Lie algebra extensions sense.}

The purpose of this paper is to show that all algebras underlying the
above actions can be obtained via the Lie algebra expansion
of the relativistic Poincar\'e algebra. The Lie algebra expansion
method was first used in \cite{Hatsuda:2001pp} and later studied from
a more general perspective and applied to 3D Chern Simons supergravity
and the M-theory superalgebra in
\cite{deAzcarraga:2002xi,deAzcarraga:2007et}.\footnote{See also
  \cite{Izaurieta:2006zz} for a generalization that
  uses semigroups and \cite{deAzcarraga:2011pa} for
  other applications of the method. The expansion method may also lead to Maxwell (super-)algebras
 \cite{deAzcarraga:2012zv,Concha:2014xfa} and even non-relativistic Maxwell algebras \cite{Aviles:2018jzw}.
 For an even more general expansion method, based of free Lie algebras, see \cite{Gomis:2017cmt}. This leads to even bigger finite-dimensional algebras which can be truncated to the ones that follow from a Lie algebra expansion.
 }
 The Lie algebra expansion method
is a general method that in the present case allows one to obtain from
the relativistic Poincar\'e algebra a number of extensions of the
non-relativistic $(D,p)$ Galilei algebra containing extra generators
some of which can be central extensions.  The first algebra in the
expansion is always the un-extended Galilei algebra which has the
additional feature that it can be obtained as a
In\"on\"u-Wigner contraction of the Poincar\'e algebra.  The other
algebras are obtained as consistent truncations of the full expansion.

This method can also be used to obtain actions for the
non-relativistic gravity theories associated to the finite
extended-Galilei algebras obtained as truncations. In most cases, but,
unfortunately, not always, it produces an invariant action.  In
particular, it gives an invariant action for the $(D,p)$ Galilei
algebras, for any $D$ and $p$. These actions can be obtained by taking
a non-relativistic limit of the Einstein-Hilbert action that mimics
the corresponding In\"on\"u-Wigner contraction of the Poincar\'e
algebra \cite{Bergshoeff:2017btm}.  Recently, by applying the Lie
algebra expansion directly at the level of the action, it has been
observed that in three dimensions the first algebra that occurs in the
expansion beyond the Galilei algebra, also allows an action
\cite{Izquierdo}. In fact, it precisely leads to the action of
\cite{Bergshoeff:2016lwr,Hartong:2016yrf}. Even the supersymmetric
action of \cite{Bergshoeff:2016lwr} was derived in this way.

In this paper we are going to explain how the Lie algebra expansion
procedure works, applying it to a few important examples. In
particular, we will derive a rule that tells, for a given truncation
and given values of $D$ and $p$, how many orders one needs to go in
the expansion beyond the lowest-order (Galilei algebra) for an action
to exist. For instance, in $D=4$ we will find that in the string case
the first algebra beyond the string Galilei algebra leads to the
action of \cite{Bergshoeff:2018vfn} but that, in the particle case,
one must take the second algebra in the expansion, which is precisely
the algebra of \cite{Hansen:2018ofj}.

The organization of this paper is as follows. In section 2 we will
give a formal description of the Lie algebra expansion
procedure. Next, in section 3, we will give explicit expressions for
the Lie algebra expansion of the Poincar\'e algebra for general $D$
and $p$.  In section 4, we will discuss a few subtleties that occur
when constructing an invariant action.  In particular, we will discuss
the role of the so-called $P$-symmetries and we will give a rule that
states for which truncations and which values of $D$ and $p$ an
invariant action exist. In the next section, we will give five
examples of actions for non-relativistic gravity thereby re-producing
several results given in the literature. Finally, in the conclusions,
we will discuss a relation between the Lie algebra expansion and
taking non-relativistic limits. We will also mention how the Lie
algebra expansion procedure can be applied to a variety of new
situations.  Finally, we have delegated the full proof of the rule
that indicates when an invariant action exists to appendix A while in
appendix B we give a few more details about the third example in the
main text.

\section{Lie algebra expansions}

The Lie algebra expansion is a method that, starting from a given Lie
algebra, leads to new Lie algebras.  These algebras are, in general,
of higher dimension than the original one. In this section we give a
general description of the method.

Quite generally, the Lie algebra expansion of an algebra is obtained
by formally expanding the Maurer-Cartan one-forms of the dual version
of the algebra in a power series of a parameter $\lambda$, each
coefficient being a new one-form. One next substitutes
this expansion into the Maurer-Cartan equations of the algebra and
identifies equal powers of $\lambda$.  This results into an infinite
set of Maurer-Cartan equations that can later be truncated
consistently to obtain a number of finite-dimensional Lie algebras.

To be specific, we describe the method in the case that is relevant to
us, that is, when the original algebra $\mathfrak{g}$ can be split, as
a vector space, into the direct sum $\mathfrak{g}=V_{0}\oplus V_{1}$
of two subspaces $V_{0}$ and $V_{1}$ with
\begin{equation}
  \label{2.1}
  [V_{0},V_{0}] \subset V_{0} \ ,
  \quad
  [V_{0},V_{1}] \subset V_{1} \ ,
  \quad [V_{1},V_{1}] \subset V_{0}\, .
\end{equation}
This means that $V_{0}$ and $V_{1}$ form a symmetric
pair.\footnote{For a discussion of other cases, see
  \cite{deAzcarraga:2002xi}.} Let $X_{i}$,
$i=1\dots \textrm{dim} \mathfrak{g}$, be a set of generators of the
starting Lie algebra $\mathfrak{g}$ whose commutation rules read
\begin{equation}\label{2.2}
  [X_{i} ,X_{j}] = C_{ij}{}^{k} X_{k}\,.
\end{equation}
Let $X_{i} = \{ X_{i_{0}}, X_{i_{1}} \}$ with $X_{i_{0}}\in V_{0}$,
$X_{i_{1}}\in V_{1}$, $i_{0}= 1\dots \textrm{dim} V_{0}$,
$i_{1}= 1\dots \textrm{dim} V_{1}$. Then, the condition \eqref{2.1}
implies that
\begin{equation}
  C_{i_{0}j_{0}}{}^{k_{1}}=0\,,
  \hskip 1truecm
  C_{i_{0}j_{1}}{}^{k_{0}}=0\,,
  \hskip 1truecm
C_{i_{1}j_{1}}{}^{k_{1}}=0\,.
\end{equation}

We now consider the dual formulation in terms of the Maurer-Cartan
one-forms $\omega^{i}=\{\omega^{i_{0}}, \omega^{i_{1}} \}$. The
commutators \eqref{2.2} are equivalent to the Maurer-Cartan equations
\begin{equation}
  \label{2.3}
     d \omega^{k} = - \frac{1}{2} C_{ij}{}^{k} \omega^{i} \wedge \omega^{j} \ .
\end{equation}
The Jacobi identity in this formulation corresponds to   $dd \equiv 0$.
It is now consistent to expand the Maurer-Cartan one-forms as follows:
\begin{subequations}
\begin{align}
  \omega^{i_{0}}
  &=
    \sum_{\alpha_{0}=0, \, \alpha_{0}\,\, \textrm{even}}^{\infty}
    \accentset{(\alpha_{0})}{\omega}{}^{\, i_{0}}
    \lambda^{\alpha_{0}} \ ,
  \\
  \nonumber \\
  \omega^{\, i_{1}}
  &=
    \sum_{\alpha_{1}=1, \, \alpha_{1}\,\, \textrm{odd}}^{\infty}
    \accentset{(\alpha_{1})}{\omega}{}^{\, i_{1}} \lambda^{\alpha_{1}} \ ,
\end{align}  \label{2.4}
\end{subequations}
where we have introduced new indices $(\alpha_{0})$ and $(\alpha_{1})$
on top of the Maurer-Cartan one-forms to indicate the order of the
corresponding terms in the expansion.  By inserting these expansions
into the Maurer-Cartan equations \eqref{2.3} and identifying equal
powers of $\lambda$, the following infinite set of Maurer-Cartan
equations is obtained:
\begin{equation}
  \label{2.5}
  d \accentset{(\gamma_{s})}{\omega}{}^{\, k_{s}}
  =
  - \frac{1}{2} C_{i_{p},\alpha_{p}\, j_{q},\beta_{q}}{}^{k_{s},\gamma_{s}}
  \accentset{(\alpha_{p})}{\omega}{}^{\, i_{p}}\wedge
  \accentset{(\beta_{q})}{\omega}{}^{\, j_{q}} \ ,
\end{equation}
where $s,p,q=0,1$ and
\begin{equation}
  \label{2.5a}
  C_{i_{p},\alpha_{p}\, j_{q},\beta_{q}}{}^{k_{s},\gamma_{s}}
  =
  \left\{
    \begin{array}{ll}
      C_{i_{p}\, j_{q}}{}^{k_{s}}\, ,
      &
        \,\,\,\,\hbox{$\gamma_{s}=\alpha_{p} +  \beta_{q}$;} \\
      & \\
      0\, , & \,\,\,\,\hbox{otherwise.} \\
    \end{array}
  \right.
\end{equation}
These equations are dual to an infinite set of commutators that
satisfy the Jacobi identity.

To obtain finite-dimensional Lie algebras, we substitute  the following
finite expansions, for given integers $(N_{0},N_{1})$,
\begin{subequations}\label{2.6}
\begin{align}
  \omega^{i_{0}}
  &=
    \sum_{\alpha_{0}=0, \, \alpha_{0}\,\, \textrm{even}}^{N_{0}}
    \accentset{(\alpha_{0})}{\omega}{}^{\, i_{0}} \lambda^{\alpha_{0}}\ ,
  \\
  \nonumber \\
  \omega^{i_{1}}
  &=
    \sum_{\alpha_{1}=1, \, \alpha_{1}\,\, \textrm{odd}}^{N_{1}}
    \accentset{(\alpha_{1})}{\omega}{}^{\, i_{1}} \lambda^{\alpha_{1}} \,,
\end{align}
\end{subequations}
in such a way that the result is a consistent set of Maurer-Cartan
equations. This is true if and only if one of the following conditions
is satisfied \cite{deAzcarraga:2002xi}:
\begin{equation}
  \label{2.7}
  N_{0}= N_{1}+1\hskip .5truecm \,\,\,\,
  \text{or}\hskip .5truecm \,\,\,\,
  N_{1}= N_{0} +1 \ .
\end{equation}

We will denote the algebras corresponding to these two conditions as
$\mathfrak{g}(N_{0},N_{1})=\mathfrak{g}(N+1,N)$ and
$\mathfrak{g}(N_{0},N_{1})=\mathfrak{g}(N, N+1)$, respectively. Note
that in the first case $N$ has to be odd, while in the second it has
to be even

We may also consider the gauge fields
$\accentset{(\alpha_{s})}{A}{}^{\, i_{s}}$ and curvatures
$\accentset{(\alpha_{s})}{F}{}^{\, i_{s}}$ for the new Lie
algebras. They are related by

\begin{equation}
  \label{2.8}
  \accentset{(\gamma_{s})}{F}{}^{\,\,\, k_{s}}
  =
  d \accentset{(\gamma_{s})}{A}{}^{\,\, k_{s}}
  + \frac{1}{2} C_{i_{p},\alpha_{p}\, j_{q},\beta_{q}}{}^{k_{s},\gamma_{s}}
  \accentset{(\alpha_{p})}{A}{}^{\, i_{p}}
  \wedge \accentset{(\beta_{q})}{A}{}^{\,j_{q}}
\end{equation}
with $\gamma_{s} = \alpha_{p}+\beta_{q}$. The gauge fields transform
as
\begin{equation}
  \label{2.9}
  \delta \accentset{(\gamma_{s})}{A}{}^{\,\, k_{s}}
  =
  d \accentset{(\gamma_{s})}{a}{}^{\,\,\, i_{s}}
  -C_{i_{p},\alpha_{p}\, j_{q},\beta_{q}}{}^{k_{s},\gamma_{s}}\,\,
  \accentset{(\alpha_{p})}{a}{}^{\,\, i_{p}}
  \accentset{(\beta_{q})}{A}{}^{\, j_{q}} \ ,
\end{equation}
where $\accentset{(\alpha_{s})}{a}{}^{\,\, i_{s}}$ is the parameter of
the gauge transformation corresponding to
$\accentset{(\alpha_{s})}{A}{}^{\, i_{s}}$.  Alternatively, these
equations may be obtained by applying the expansion method to the
gauge curvatures and variations of the original algebra
\begin{subequations}\label{2.10}
\begin{align}
  F^{k}&= dA^{k} + \frac{1}{2} C_{ij}{}^{k} A^{i} \wedge A^{j} \ ,
  \\
  \nonumber \\
 \delta A^{k} &= da^{k} - C_{ij}{}^{k} a^{i} A^{j}\ ,
\end{align}
\end{subequations}
where the gauge fields, curvatures and gauge parameters are expanded as in
\eqref{2.4} or \eqref{2.6} by simply replacing the symbol $\omega$ by $A$, $F$
or $a$. The Maurer-Cartan equations of the expanded algebras are then
recovered by simply setting the curvatures equal to zero.

\vskip.2truecm

Summarizing, by using the method of Lie algebra expansions one can
derive from a given Lie algebra a number of extended algebras with
corresponding gauge fields and curvatures. Given a Lie algebra
$\mathfrak{g}$ with corresponding gauge fields $A^{i}$ and curvatures
$F^{i}$ for $\mathfrak{g}$, one just has to follow these steps:

\vskip .3truecm

\begin{enumerate}
\item Find a symmetric splitting $\mathfrak{g}=V_{0}\oplus V_{1}$,
  \textit{i.e.}, a splitting with the structure given in
  eq.~\eqref{2.1}.
\item Expand the gauge fields and curvatures corresponding to the
  generators in $V_{0}$ (resp. $V_{1}$) in odd (resp.~even) powers of
  a parameter $\lambda$.
\item Set the curvatures equal to zero and obtain an infinite set of
  Maurer-Cartan equations that can be truncated consistently into the
  finite algebras $\mathfrak{g}(N,N+1)$ and $\mathfrak{g}(N+1,N)$.
\end{enumerate}

Finally, we stress that in the procedure defined above we always expand in terms of a parameter $\lambda$ using the dual formulation, we never expand the Lie algebra itself.
\section{The Lie algebra expansion of the Poincar\'e algebra}

In this section we apply the Lie algebra expansion method to the
specific case of the Poincar\'e algebra with the aim of constructing
actions for non-relativistic gravity in the next section.  Our
starting point is the D-dimensional Poincar\'e algebra with the
following commutation relations:
\begin{multicols}{2}
\begin{subequations}\label{eq:PoincareAlgebraCommutators}
\setlength{\abovedisplayskip}{-15pt}
\allowdisplaybreaks
\begin{align}
[P_{\hat{A}},P_{\hat{B}}]& = 0\, ,\\
    [M_{\hat{A}\hat{B}},P_{\hat{C}}]
    & = 2\eta_{\hat{C}[\hat{B}}P_{\hat{A}]}\, , \\
    [M_{\hat{A}\hat{B}},M_{\hat{C}\hat{D}}]
    &= 4\eta_{[\hat{A}[\hat{C}}M_{\hat{D}]\hat{B}]}\, ,
\end{align}
\end{subequations}
\end{multicols}
\noindent
where $P_{\hat{A}}$ and $M_{\hat{A}\hat{B}}$ are the generators of
spacetime translations and Lorentz transformations, respectively.  The
hatted indices run over $\hat{A}=0,...,D-1$ and we have chosen the
Minkowski metric to have mostly plus signature. As we have discussed
in the introduction, for a $p$-brane, it is natural to decompose the
indices as
\begin{equation}
\hat{A}=\{A,a\}\,,\hskip 1truecm A=0,1,\cdots,p;\,\quad  a=p+1,\cdots,D-1\,.
\end{equation}
This induces the following decomposition of the generators:
\begin{subequations}
\begin{align}
M_{\hat{A}\hat{B}}&\rightarrow\{J_{AB}, G_{Aa},J_{ab}\}\, ,\\
 P_{\hat{A}}&\rightarrow\{H_{A}, P_{a}\}\, .
\end{align}
\end{subequations}
The non-vanishing commutation relations of these generators are given by
\begin{multicols}{2}
\begin{subequations}
\setlength{\abovedisplayskip}{-15pt}
\allowdisplaybreaks
\begin{align}
[J_{AB},H_{C}]&= 2\eta_{C[B}H_{A]}\, , \\
[J_{ab},P_{c}]&= 2\eta_{c[b}P_{a]}\, ,\\
[J_{AB},J_{CD}]&= 4\eta_{[A[C}J_{D]B]}\, , \\
[J_{ab},J_{cd}]&=4\eta_{[a[c}J_{d]b]}\, ,\\
[G_{Aa},P_{b}]&=\delta_{ab}H_{A}\, , \\
[G_{Aa},H_{B}]&=-\eta_{AB}P_{a}\, ,\\
[J_{AB},G_{Cd}]&=2\eta_{C[B}G_{A]d}\, , \\
[J_{ab},G_{Ac}]&=-2\delta_{c[b|}G_{A|a]}\, ,\\
[G_{Aa},G_{Bb}]&=-\eta_{AB}J_{ab}-\delta_{ab}J_{AB}\, .
\end{align}
\end{subequations}
\end{multicols}

For each generator, we introduce a 1-form gauge field as follows:
\begin{multicols}{2}
\begin{subequations}
\setlength{\abovedisplayskip}{-15pt}
\allowdisplaybreaks
\begin{align}
J_{AB}&\rightarrow\ \Omega^{AB}\, ,\label{gf1} \\
J_{ab}&\rightarrow\ \Omega^{ab}\, ,\\
G_{Ab}&\rightarrow\ \Omega^{Ab}\, , \\
P_{a}&\rightarrow\ E^{a}\, ,\\
H_{A}&\rightarrow\ \tau^{A}\, .\label{gf5}
\end{align}
\end{subequations}
\end{multicols}
The curvatures for these gauge fields are given by
\begin{subequations}\label{curvatures}
\begin{align}
  R^{AB}(J)
  & =
    d\Omega^{AB}-\Omega^{AC}\wedge \Omega^{B}_{\phantom{B}C}
    -\Omega^{Aa}\wedge\Omega^{B}_{\ a}\, ,
  \\[4pt]
  R^{ab}(J)
  & =
    d\Omega^{ab}-\Omega^{ac}\wedge
    \Omega^{b}_{\phantom{b}c}-\Omega^{Aa}\wedge\Omega_{A}^{\ b}\, ,
  \\[4pt]
  R^{Aa}(G)
  & =
    d\Omega^{Aa}+\Omega^{AB}\wedge \Omega_{B}^{\phantom{B}a}
    +\Omega^{ab}\wedge\Omega^{A}_{\phantom{A}b}\, ,
  \\[4pt]
  R^{a}(P)
  & =
    dE^{a}+\Omega^{ab}\wedge E_{b}-\Omega^{Aa}\wedge \tau_{A}\, ,
  \\[4pt]
  R^{A}(H)
  & =
    d\tau^{A}+\Omega^{AB}\wedge \tau_{B}+\Omega^{Aa}\wedge E_{a}\,\quad .
\end{align}
\end{subequations}

We now consider the splitting of the Poincar\'e algebra
$\mathfrak{g}=V_{0}\oplus V_{1}$ that will induce a Lie algebra
expansion where the lowest order algebra is given by the $(D,p)$
Galilei algebra.  This splitting is defined by taking the following
choice for $V_{0}$ and $V_{1}$:
\begin{equation}
V_{0}=\{H_{A},J_{AB},J_{ab}\}\,,\hskip 2truecm
V_{1}=\{P_{a},G_{Ab}\}\,.\label{eq:splittingGalilei}
\end{equation}
We recognize that $V_{0}$ and $V_{1}$ form a symmetric pair satisfying
eq.~\eqref{2.1}. Following the Lie algebra expansion method, we expand
the gauge fields as follows:
\allowdisplaybreaks
\begin{subequations}\label{infinite}
\begin{alignat}{2}
  \tau^{A}
  & =
  \sum_{k=0,\,\, k\,\, \text{even}}^{N_{0}}\lambda^{k}\
  \accentset{(k)}{\tau}^{A}
  &&
  =\accentset{(0)}{\tau}^{A}+\lambda^{2}\
  \accentset{(2)}{\tau}^{A}+\ldots\, ,
  \\[4pt]
  \Omega^{AB}
  & =
  \sum_{k=0,\,\, k\,\, \text{even}}^{N_{0}}\lambda^{k}\
  \accentset{(k)}{\Omega}^{AB}
  &&
  =\accentset{(0)}{\Omega}^{AB}+\lambda^{2}\
  \accentset{(2)}{\Omega}^{AB}+\ldots\, ,
  \\[4pt]
  \Omega^{ab}
  &
  = \sum_{k=0,\,\,k\,\, \text{even}}^{N_{0}}\lambda^{k}\
  \accentset{(k)}{\Omega}^{ab}
  &&
  =\accentset{(0)}{\Omega}^{ab}+\lambda^{2}\
  \accentset{(2)}{\Omega}^{ab}+\ldots\, ,
  \\[4pt]
  \Omega^{Ab}
  & =
  \sum_{k=1,\,\,k\,\, \text{odd}}^{N_{1}}\lambda^{k}\ \accentset{(k)}{\Omega}^{Ab}
  &&
  =\lambda\ \accentset{(1)}{\Omega}^{Ab}+\ldots\, ,
  \\[4pt]
  E^{a}
  & =
  \sum_{k=1,\,\,k\,\, \text{odd}}^{N_{1}}\lambda^{k}\
  \accentset{(k)}{E}^{a}
  &&
  =\lambda\ \accentset{(1)}{E}^{a}+\ldots\, .
\end{alignat}
\end{subequations}

According to the general method, the infinite expansion in
eqs.~\eqref{infinite} can be consistently truncated by taking one of
the two conditions $N_{0}=N_{1}\pm 1$.  The truncation of the algebras
induces similar truncations of the expansions of the curvatures and
transformation rules for the finite-dimensional algebras.  For
instance, the curvatures corresponding to the finite-dimensional
algebra $\mathfrak{g}(1,2)$ are given by\footnote{Note that in the
  expression of the curvatures (using the notation of
  \cite{deAzcarraga:2002xi})
\begin{equation}
F^{k_{s},\alpha_{s}}=dA^{k_{s},\alpha_{s}}+\frac{1}{2}C_{i_{p},\beta_{p}\, j_{q},\gamma_{q}}{}^{k_{s},\alpha_{s}}
                A^{i_{p},\beta_{p}}\wedge A^{j_{q},\beta_{q}}
\end{equation}
the structure constants are zero when $\alpha_{s}> \max\{N_{0},N_{1}\}$.}
\allowdisplaybreaks
\begin{subequations}
\begin{align}
  \accentset{(0)}{R}^{AB}(J)
  &
    = d\accentset{(0)}{\Omega}^{AB}-\accentset{(0)}{\Omega^{AC}}\wedge
    \accentset{(0)}{\Omega}^{B}_{\phantom{B}C}\, ,
  \\[4pt]
  \accentset{(2)}{R}^{AB}(J)
  &
    = d\accentset{(2)}{\Omega}^{AB}-
\accentset{(2)}{\Omega}^{AC}\wedge
    \accentset{(0)}{\Omega}^{B}_{\phantom{B}C}-\accentset{(0)}{\Omega}^{AC}\wedge
    \accentset{(2)}{\Omega}^{B}_{\phantom{B}C}-\accentset{(1)}{\Omega}^{Aa}\wedge\accentset{(1)}{\Omega}^{B}_{\
    a}\, ,
  \\[4pt]
  \accentset{(0)}{R}^{ab}(J)
  &
    = d\accentset{(0)}{\Omega}^{ab}-\accentset{(0)}{\Omega^{ac}}\wedge
    \accentset{(0)}{\Omega}^{b}_{\phantom{b}c}\, ,
  \\[4pt]
  \accentset{(2)}{R}^{ab}(J)
  &
    = d\accentset{(2)}{\Omega}^{ab}-\accentset{(2)}{\Omega}^{ac}\wedge
    \accentset{(0)}{\Omega}^{b}_{\phantom{b}c}-\accentset{(0)}{\Omega}^{ac}\wedge
    \accentset{(2)}{\Omega}^{b}_{\phantom{b}c}-\accentset{(1)}{\Omega}^{Aa}
    \wedge\accentset{(1)}{\Omega}_{A}^{\ b}\, ,\\[4pt]
  \accentset{(0)}{R}^{A}(H)
  &
    = d\accentset{(0)}{\tau}^{A}+\accentset{(0)}{\Omega}^{AB}\wedge
    \accentset{(0)}{\tau}_{B}\, ,
  \\[4pt]
  \accentset{(2)}{R}^{A}(H)
  &
    = d\accentset{(2)}{\tau}^{A}+\accentset{(2)}{\Omega}^{AB}\wedge \accentset{(0)}{\tau}_{B}
    +\accentset{(0)}{\Omega}^{AB}\wedge
    \accentset{(2)}{\tau}_{B}+\accentset{(1)}{\Omega}^{Aa}\wedge
    \accentset{(1)}{E}_{a}\, ,
  \\[4pt]
  \accentset{(1)}{R}^{Aa}(G)
  &
    = d\accentset{(1)}{\Omega}^{Aa}
    +\accentset{(0)}{\Omega}^{AB}\wedge
    \accentset{(1)}{\Omega}_{B}^{\phantom{B}a}+\accentset{(0)}{\Omega}^{ab}
    \wedge\accentset{(1)}{\Omega}^{A}_{\phantom{B}b}\, ,
  \\[4pt]
  \accentset{(1)}{R}^{a}(P)
  &
    = d\accentset{(1)}{E}^{a}
    +\accentset{(0)}{\Omega}^{ab}\wedge \accentset{(1)}{E}_{b}
    -\accentset{(1)}{\Omega}^{Aa}\wedge \accentset{(0)}{\tau}_{A}\, .
\end{align}
\end{subequations}

This concludes our general discussion of the expansion and consistent
finite-di\-men\-sio\-nal truncations of the Poincar\'e algebra and the
associated gauge fields and curvatures using the method explained
before. It gives an infinite number of finite-dimensional algebras and
associated gauge fields and curvatures, but a dynamical principle
(such as the extremization of an action) needs to be formulated in
order to construct physical non-relativistic theories of gravity with
them. If we use an action principle, the action will have to be
invariant under the symmetries that we want the physical theory to
inherit.

The method of Lie algebra expansions gives a simple recipe to
construct invariant actions that we are going to review in the next
section and that we are going to apply to the present case. For some
of the subalgebras generated by this method, though, it does not
give an invariant action.

\section{Actions}

\subsection{The Expanded Action}

We may construct actions as follows.  Let $B$ be an invariant $D$-form
defined on the algebra $\mathfrak{g}$ generated by the gauge fields
$A^{i}$ and curvatures $F^{i}$ corresponding to $\mathfrak{g}$.  We
assume that these gauge fields and curvatures are realized on a
spacetime manifold $\mathcal{M}$, i.e.,
$A^{i}= A^{i}{}_{\mu} dx^{\mu}$ and
$F^{i} = F^{i}{}_{\mu\nu} dx^{\mu} \wedge dx^{\nu}$. We may now use
$B$ to define an action
\begin{equation}\label{g}
I[A]=\int_{\mathcal{M}} B\,,
\end{equation}
that is invariant under the gauge transformations of some subalgebra
and the diffeomorphisms of $\mathcal{M}$.

We now substitute $F^{i}$ and $A^{i}$ into the action \eqref{g} by
their infinite expansions, which results into the following expansion
of $B$ and the action $I[A]$:
\begin{equation}
\label{2.11}
I[A](\lambda)
=
\sum_{k=0}^{\infty} \accentset{(k)}{I} \lambda^{k}
=
\sum_{k=0}^{\infty} \left(\int_{\mathcal{M}} \accentset{(k)}{B}\right) \lambda^{k}\ .
\end{equation}

If the original action is invariant under the gauge transformation for
a certain gauge parameter, say $a^{i_{0}}$,
\begin{equation}
\delta A^{i} = \delta^{i}{}_{i_{0}} da^{i_{0}} - C_{i_{0} \, j}{}^{i} a^{i_{0}} A^{j}\,,
\end{equation}
then, by expanding the l.h.s.~of the equation $\delta I=0$, we see
that all the terms $\accentset{(k)}{I}$ are invariant under the
transformations with gauge parameters
$\accentset{(\alpha)}a{}^{\,\, i_{0}}$ that appear in its expansion,
provided we keep the expansion infinite.

Consider now the term $\accentset{(k)}{I}$ appearing at a certain
order $k$ in the expansion of the action.  This action term will
contain only a finite number of fields. However, due to the truncation
of the expanded gauge algebra, it will not be necessarily
invariant. In particular, the action $\accentset{(k)}{I}$ will only be
invariant provided that the gauge algebra expansion is the smallest
truncated expansion that contains all the fields appearing in
$\accentset{(k)}{I}$. A further truncation to a smaller expansion
corresponding to smaller finite-dimensional algebra may result into an
action that is not gauge invariant under the variations of the
parameter $\accentset{(\alpha)}a{}^{\,\, i_{0}}$. Below, in the case
of the Poincar\'e algebra, we will derive a rule that states for which
algebras in the Lie algebra expansion this method provides an invariant
action.

\vskip .2truecm

Summarizing, the method of Lie algebra expansions can be used to
construct invariant actions corresponding to finite-dimensional
algebras that are extensions of the original Lie algebra
$\mathfrak{g}$ by the following steps.  Given a Lie algebra
$\mathfrak{g}$ and an action invariant under some subalgebra defined
as $I=\int_{M}B$ where $B$ is the exterior product of gauge fields and
curvatures $A^{i}$, $F^{i}$ for $\mathfrak{g}$,

\vskip .3truecm

\begin{enumerate}
\item Insert the infinite expansion of $A^{i}$ and $F^{i}$ into the
  original action $\int_{M}B$ and obtain an infinite set of terms
  $\accentset{(k)}{I}$.
\item In order for $\accentset{(k)}{I}$ to define an invariant action,
  select the truncated gauge algebra to be such that it is the
  smallest algebra whose fields give rise to {\sl all} terms in
  $\accentset{(k)}{I}$ that occurred in the infinite expansion.
\end{enumerate}

\vskip .2truecm

\subsection{Einstein-Hilbert Action}

We now apply the general procedure explained in the previous
subsection to the Poincar\'e algebra and the first-order
Einstein-Hilbert action, which, as it is well known, is only invariant
under general coordinate transformations and the Lorentz
transformations of the Poincar\'e algebra but not under the
$P$-transformations that are generated by the $P_{\hat A}$ generators of the same Poincar\'e
algebra. These transformations require a separate discussion.

\subsubsection{$P$-Transformations}

To explain our point it is sufficient to consider the 4D case. The 4D
Einstein-Hilbert (EH) action in first-order formulation is given by
\begin{equation}\label{standard}
  S_{\rm EH}
  =
  \int d^{4}x \,E E^{\mu}{}_{\hat{A}} E^{\nu}{}_{\hat{B}}\,
  R_{\mu\nu}{}^{\hat{A} \hat{B}}(M)\,,
\end{equation}
where $E^{\mu}{}_{\hat{A}}$ is the inverse Vierbein and
$E= \textrm{det}\, E_{\mu}{}^{\hat{A}}$.  This action is invariant under
general coordinate transformations, with pa\-ra\-meter $\xi^{\mu}$,
and under the Lorentz transformations of the Poincar\'e algebra with
pa\-ra\-meters $\Lambda^{\hat{A}\hat{B}}$:
\begin{subequations}
\begin{eqnarray}\label{gctL}
  \delta E_{\mu}{}^{\hat{A}}
  & = &
        \xi^{\mu}\partial_{\mu} E_{\mu}{}^{\hat{A}}
        +\partial_{\mu}\xi^{\lambda} E_{\lambda} {}^{\hat{A}}
        +\Lambda^{\hat{A}}{}_{\hat{B}}E_{\mu}{}^{\hat{B}}\, ,
  \\[4pt]
  \delta \Omega_{\mu}{}^{\hat{A}\hat{B}}
  & = &
        \xi^{\lambda}\partial_{\lambda}  \Omega_{\mu}{}^{\hat{A}\hat{B}}
        +\partial_{\mu}\xi^{\lambda}  \Omega_{\lambda}{}^{\hat{A}\hat{B}}
        +\partial_{\mu} \Lambda^{\hat{A}\hat{B}}
        +2\Lambda^{\hat{C} [\hat{A}}\Omega_{\mu}{}^{\hat{B}]}{}_{\hat{C}}\,.
\end{eqnarray}
\end{subequations}

The action \eqref{standard} is, however, not invariant under the
$P$-transformations, with parameters $\eta^{\hat{A}}$, of the Poincar\'e
algebra:
\begin{equation}\label{P}
  \delta E_{\mu}{}^{\hat{A}}
  =
  \partial_{\mu} \eta^{\hat{A}} -\Omega_{\mu}{}^{\hat{A}}{}_{\hat{B}}\eta^{\hat{B}} \,.
\end{equation}
Instead, the action transforms into a term proportional to the
equation of motion of the spin connection
$R_{\mu\nu}{}^{\hat{A}}(P)=0$. Such a variation can always be
cancelled by a modification of the $P$-transformation rule of the spin
connection that consists in the addition of curvature  terms that do not follow from the original Poincar\'e algebra.

In order to find those terms and the modified $P$-transformation rules
we start by observing that the first-order action \eqref{standard} is
also invariant under the following so-called `trivial' symmetry with
parameter $\sigma^{\hat{A}}$:\footnote{ Trivial symmetries, also
  called `equation of motion' symmetries, have the distinguishing
  feature that all terms in the transformation rules are proportional
  to the equations of motion. This implies that trivial symmetries
  correspond to vanishing Noether charges.}
\begin{subequations}
\begin{eqnarray}\label{eom}
  \delta E_{\mu}{}^{\hat{A}}
  & = &
        R_{\mu\nu}{}^{\hat{A}}(P) \sigma^{\nu}\,,
  \\[4pt]
  \delta \Omega_{\mu}{}^{\hat{A}\hat{B}}
  & = &
        2R_{\mu}{}^{[\hat{A}}(M) \sigma^{\hat{B}]}
        +  E_{\mu}{}^{[\hat{A}}R_{\hat{C}}{}^{\hat{B}]}(M) \sigma^{\hat{C}}
        -\frac{1}{2}  E_{\mu}{}^{[\hat{A}} R(M) \sigma^{\hat{B}]}\,,
\end{eqnarray}
\end{subequations}
with $R_\mu{}^{\hat A}(M) = E^\nu{}_{\hat B} R_{\mu\nu}{}^{\hat A\hat B}(M), R(M) = E^\mu{}_{\hat A}R_\mu{}^{\hat A}(M)$
and
$\sigma^{\nu} \equiv \sigma^{\hat{B}} E^{\nu}{}_{\hat{B}}$.

It turns out that the $P$-transformation of the Vierbein with
parameter $\eta^{\hat{A}}$ in eq.~(\ref{P}) can be written as the sum
of a general coordinate transformation, a Lorentz transformation and a
trivial symmetry of the kind introduced above with parameters
\begin{equation}
  \xi^{\mu}
  = \eta^{\mu}\,,
  \hskip 1truecm
  \Lambda^{\hat{A}\hat{B}}=\eta^{\lambda}\Omega_{\lambda}{}^{\hat{A}\hat{B}}\,,
  \hskip 1truecm
  \sigma^{\hat{A}} =  \eta^{\hat{A}}\,,
\end{equation}
with $\eta^{\mu} \equiv \eta^{\hat{B}} E^{\mu}{}_{\hat{B}}$.

Therefore, we can reinterpret a $P$-transformation, not a new symmetry
of the action \eqref{standard}, but just as a linear combination of
three symmetries of the action: a general coordinate transformation, a
Lorentz transformation and a trivial symmetry. Since the same linear
combination must be realized on the spin connection, we conclude that
the $P$-transformation of the Vierbein and spin connection that leave
the action invariant are given by
\begin{subequations}
\begin{eqnarray}\label{Ptransf}
  \delta_\eta  E_{\mu}{}^{\hat{A}}
  & = &
        D_{\mu} (\Omega)\eta^{\hat{A}}\,,
  \\[4pt]
  \delta_\eta \Omega_{\mu}{}^{\hat{A}\hat{B}}
  & = &
        \eta^{\lambda} R_{\lambda\mu}{}^{\hat{A}\hat{B}}(M)
        +2R_{\mu}{}^{[\hat{A}}(M) \eta^{\hat{B}]}
        +E_{\mu}{}^{[\hat{A}}R_{\hat{C}}{}^{\hat{B}]}(M) \eta^{\hat{C}}
        -\frac{1}{2}E_{\mu}{}^{[\hat{A}} R(M) \eta^{\hat{B}]}\,.
\end{eqnarray}
\end{subequations}

\vskip.2truecm

Summarizing, when considering the first-order Einstein-Hilbert action
\eqref{standard} and its Lie algebra expansion, see below, we should
not consider the general coordinate transformations and
$P$-transformations as two independent symmetries.\footnote{Observe
  that, if they were independent and both of them were realized in the
  action together with the lorentz transformations, the theory would
  have no propagating degrees of freedom left.} Both have their
advantages. On the one hand, it is much easier to expand the general
coordinate transformations but, on the other hand, the
$P$-transformations are more directly related to the underlying
algebra. In the following we will consider the general coordinate
transformations in the Lie algebra expansion but we will
explain how, after the Lie algebra expansion, the expanded general
coordinate transformations remain related, up to trivial symmetries,
to the expanded $P$-transformations.

\subsubsection{Conditions for Gauge Invariance of the Action}

When discussing the Lie algebra expansion and invariant actions, it is
convenient to use an alternative expression for the EH action
\eqref{standard} given by
\begin{equation}\label{I1}
  S_{\rm EH}
  =
  \frac{1}{4}\int d^{4}x\, \epsilon^{\mu\nu\rho\sigma}
  \epsilon_{\hat{A}\hat{B}\hat{C}\hat{D}}
  E_{\mu}{}^{\hat{A}}E_\nu{}^{\hat{B}} R_{\rho\sigma}{}^{\hat{C}\hat{D}}(M)\ .
\end{equation}
that has the distinguishing feature that it does not contain inverse
Vierbeine. This expression can easily be derived from the more
conventional eq.~\eqref{standard} by using the standard formula for
the Vierbein determinant $E$:
\begin{equation}
  E
  \equiv
  \textrm{det}\, E_{\mu}{}^{\hat{A}}
  = \frac{1}{4!}\epsilon^{\mu\nu\rho\sigma}\epsilon_{\hat{A}\hat{B}\hat{C}\hat{D}}\,
  E_{\mu}{}^{\hat{A}}E_\nu{}^{\hat{B}}E_\rho{}^{\hat{C}}E_\sigma{}^{\hat D}\, .
\end{equation}

We now proceed to discuss invariant actions and, in particular, to
address the issue that not every term in the expansion of the
D-dimensional EH Lagrangian density
\begin{equation}\label{3.1}
  B
  =
  \epsilon_{\hat{A}_{0} \ldots \hat{A}_{D-1}}
  R^{\hat{A}_{0}\hat{A}_{1}}(M) \wedge E^{\hat{A}_2}
  \wedge \ldots \wedge E^{\hat{A}_{D-1}}\, ,
\end{equation}
leads to an invariant action. To make our point, we consider the case
$(D,p)=(4,0)$ and only the terms in the expansion that are
proportional to $R^{ab}(J)$. The terms of lowest order in the
expansion obtained in this way are of order $1$ and read
\begin{equation}\label{lowestorder}
 \accentset{(1)}{B}
  =
    -2\epsilon_{abc}\accentset{(0)}{R}^{ab}(J)\wedge\accentset{(1)}{E}^{c}\wedge
 \accentset{(0)}{\tau}\, .
\end{equation}
This term defines the lowest-order Galilei action corresponding to the
algebra $\mathfrak{g}(0,1)$.

In the un-truncated expansion the next order terms involving
$R^{ab}(J)$ occur at order $n=3$. They can be obtained from the
lowest-order term \eqref{lowestorder} by either replacing one of the
$V_{0}$ fields by a next-order one, \textit{i.e.},
$\accentset{(0)}{\Omega}_{\mu}{}^{\, ab}\ \rightarrow\
\accentset{(2)}{\Omega}_{\mu}{}^{\, ab}$ or
$\accentset{(0)}{\tau}_{\mu}\ \rightarrow
\accentset{(2)}{\tau}_{\mu}$, or by replacing the $V_{1}$ field by a
next order one, \textit{i.e.},
$\accentset{(1)}{E}_{\mu}{}^{\, a}\ \rightarrow
\accentset{(3)}{E}_{\mu}{}^{\, a}$. In the un-truncated expansion
there are no more terms contributing to $\accentset{(3)}{B}$. In other
words, there are no other replacements in $\accentset{(1)}{B}$,
involving higher-order expansion terms, that would also contribute to
$\accentset{(3)}{B}$.

All the above replacements lead to terms that occur in the  un-truncated expansion and, furthermore,  all these terms are needed  to
obtain an invariant action at order 3. However, not all terms will  survive a specific truncation
specified by $(N_{0},N_{1})$.  For instance, the truncation
$(N_{0},N_{1}) = (2,1)$ leading to the first algebra beyond the
Galilei algebra, does not contain the
$\accentset{(3)}{E}_{\mu}{}^{\, a}$ field. This implies that, although
the $\accentset{(3)}{B}$ term corresponding to the $\mathfrak{g}(2,1)$
algebra will contain all fields of that algebra, the corresponding
action will not be invariant due to the missing term involving the
$V_{1}$ field $\accentset{(3)}{E}_{\mu}{}^{\, a}$.  It is only if one
goes to the next-order algebra $\mathfrak{g}(2,3)$ that the the term in $\accentset{(3)}{B}$ containing the
$V_{1}$
field $\accentset{(3)}{E}_{\mu}{}^{\, a}$ is
included. Correspondingly, the $\accentset{(3)}{B}$ action
corresponding to this bigger algebra will contain all fields that
would also have occurred in the un-truncated expansion. It therefore
defines an invariant action.

The above argument can be refined for arbitrary values of
$(D,p;N_{0},N_{1})$. In Appendix~\ref{app-invarianceconditions} we
derive a general invariance condition that states, for given values of
$(D,p; N_{0},N_{1})$, for which $n$ the Lagrangian densities
$\accentset{(n)}{B}$ corresponding to the algebra
$\mathfrak{g}(N_{0},N_{1})$ defines an invariant action. Depending on
the values of $p$ we find that these invariance conditions read as
follows:
\begin{subequations}\label{invcond}
\begin{align}
&p=0&&\left\{
\begin{array}{lcl}
D=3 &&n \leqslant N_{0}\\ [0.2cm]
D\neq 3 &&n\leqslant N_{1}+D-4
\end{array}\right.\\
&\nonumber\\
&1 \leqslant p\leqslant D-3&&\left\{
\begin{array}{lcl}
D=p+3 &&n \leqslant N_{0}\\ [0.2cm]
D\neq p+3 &&n\leqslant N_{1}+D-p-4\\
\end{array}\right.\\
&\nonumber\\
&p=D-2&&n\leqslant N_{1}\\
&\nonumber\\
&p=D-1&&n\leqslant N_{0}
\end{align}
\end{subequations}

From these conditions, we may also derive that, for given $(D,p)$, the
smallest algebra consistent with the invariance of the Lagrangian
density $\accentset{(n)}B$ is given by
\begin{subequations}\label{invcond2}
\begin{align}
&D>p+3&& \mathfrak{g}(n-D+p+3, n-D+p+4)\\[4pt]
&D= p+3 &&\mathfrak{g}(n,n-1)\\[4pt]
&D= p+2 &&\mathfrak{g}(n-1,n)\\[4pt]
&D=p+1&&\mathfrak{g}(n,n-1)
\end{align}
\end{subequations}

In the case with $n=0$ it is understood that the algebra
$\mathfrak{g}(0,-1)$ corresponds to $V_{0}$. Note also that in the
expansion of $B$, the value of $n$ in all terms $\accentset{(n)} B$ is
even (resp.~odd) when $D-p-1$ is even (resp.~odd).

We are now in a position that allows us to perform the Lie algebra
expansion of the first-order Einstein-Hilbert Lagrangian
\eqref{3.1}. In terms of the gauge fields given in eqs.~\eqref{gf1}-\eqref{gf5}
and the curvatures defined by \eqref{curvatures} this Lagrangian
density is given by
\begin{align}
  \label{eq:D-dimexpandedLagrangian}
B & = \epsilon_{A_{0}A_{1}\ldots A_{p}a_{p+1}\ldots a_{D-1}}\left[
\binom{D-2}{p-1}R^{A_{0} A_{1}}(J)\wedge E^{a_{p+1}}\wedge
    E^{a_{p+2}}\right.
    \nonumber\\[4pt]
  &+\binom{D-2}{p+1}R^{a_{p+1} a_{p+2}}(J)\wedge \tau^{A_{0}}\wedge
    \tau^{A_{1}}
     \nonumber\\[4pt]
&\left.+2\binom{D-2}{p}R^{A_{0} a_{p+1}}(G)\wedge E^{a_{p+2}}\wedge \tau^{A_{1}}
\right]\wedge \tau^{A_{2}}\wedge \ldots \wedge \tau^{A_{p}}
\wedge E^{a_{p+3}}\wedge \ldots \wedge E^{a_{D-1}}\, .
\end{align}
Following the general procedure, considering the decomposition
\eqref{eq:splittingGalilei}, as an example we expand this Lagrangian
density order by order in $\lambda$ considering the algebra
$\mathfrak{g}(2,1)$. For general values of $D$ and $p$ the
lowest-order term in the expansion is given by
\begin{subequations}
\begin{align}
  \accentset{(D-p-3)}{B}
  =
  \epsilon_{A_{0} A_{1}\ldots A_{p}a_{p+1}\ldots a_{D-1}}
  \binom{D-2}{p+1}
  &
    \accentset{(0)}{R}^{a_{p+1} a_{p+2}}(J)\wedge
                \accentset{(0)}{\tau}^{A_{0}}\wedge \ldots \wedge
                \accentset{(0)}{\tau}^{A_{p}}\wedge
                \accentset{(1)}{E}^{a_{p+3}}\wedge \ldots \wedge
                \accentset{(1)}{E}^{a_{D-1}}\, .
\end{align}
\end{subequations}
Using the   short-hand notation
\begin{subequations}
\begin{align}
  \accentset{(k)}{E}^{a_{1}\ldots a_{k}}
  &
    =\accentset{(1)}{E}^{a_{1}}\wedge \ldots \wedge \accentset{(1)}{E}^{a_{k}}\, ,
  \\[4pt]
  \accentset{(2k)}{\tau}^{A_{1}\ldots A_{n}}
  &
    =\accentset{(2)}{\tau}^{A_{1}}\wedge \ldots .\wedge
    \accentset{(2)}{\tau}^{A_{k}}\wedge \accentset{(0)}{\tau}^{A_{k+1}}\wedge
    \ldots \wedge \accentset{(0)}{\tau}^{A_{n}}\, ,
\end{align}
\end{subequations}
we can write the term of order $D-p-3+2k$ as follows:
\begin{equation}
\begin{aligned}
  \accentset{(D-p-3+2k)}{B}\
  &
  = \left\{
    \binom{D-2}{p-1}
    \left[
      \binom{p-1}{k-1}\accentset{(0)}{R}^{A_{0}A_{1}}(J)\wedge\
      \accentset{(2k-2)}{\tau}^{\phantom{Ai}
  A_{2}\ldots A_{p}}
\right.\right.
\\
& \\
&
\left.
  +\binom{p-1}{k-2}\accentset{(2)}{R}^{A_{0}A_{1}}(J)\wedge\
  \accentset{(2k-4)}{\tau}^{\phantom{Ai}
A_{2}\ldots A_{p}}
\right]
\wedge \accentset{(D-p-1)}{E}^{a_{p+1}\ldots a_{D-1}}
\\
&\\
&
+\binom{D-2}{p+1}
\left[
  \binom{p+1}{k}\accentset{(0)}{R}^{a_{p+1}a_{p+2}}(J)\wedge
  \accentset{(2k)}{\tau}^{\phantom{A}
  A_{0}\ldots A_{p}}
\right.
\\
&  \\
&
\left.
  +\binom{p+1}{k-1}\accentset{(2)}{R}^{a_{p+1}a_{p+2}}(J)\wedge\
  \accentset{(2k-2)}{\tau}^{\phantom{Ai}\
A_{0}\ldots A_{p}}
\right]
\wedge \accentset{(D-p-3)}{E}^{a_{p+3}\ldots a_{D-1}}
\\
&\\
&
\left.
  +2(-1)^{p}\binom{D-2}{p}\binom{p}{k-1}\accentset{(1)}{R}^{A_{0}a_{p+1}}(G)\wedge\
  \accentset{(2k-2)}{\tau}^{\phantom{Ai}
A_{1}\ldots A_{p}}\wedge \ \accentset{(D-p-2)}{E}^{a_{p+2}\ldots a_{D-1}}
\right\}
\epsilon_{A_{0}\ldots A_{p}a_{p+1}\ldots a_{D-1}}\, .
\end{aligned}
\end{equation}

We note that while the lowest order term remains the same for any
$\mathfrak{g}(N_{0},N_{1})$ with $N_{1}\geqslant 2$ this is not true for
the higher-order terms.

Using the invariance conditions \eqref{invcond} it is now straightforward to
find out, in the general case, for which order $n$ the action term
$\accentset{(n)}{B}$ , for a given value of $(D,p;N_{0},N_{1})$, defines an
action that is invariant under the expanded Lorentz transformations and
general coordinate transformations.  In the next section we give five examples
thereby reproducing several results in the literature.

\section{Examples}

In this section we present five examples of invariant actions that satisfy the
invariance conditions \eqref{invcond}, see also Table~\ref{table:table2}.

\begin{description}

\item[Example 1:]\ $(D,p;N_{0},N_{1})= (4,0;0,1)$. This example leads to the
  4D Galilei gravity action which was extensively discussed in
  \cite{Bergshoeff:2017btm}. The same action can be obtained by taking a
  particular limit of the Einstein-Hilbert action \cite{Bergshoeff:2017btm}.

\item[Example 2:]\ $(D,p;N_{0},N_{1})= (3,0;2,1)$. This example reproduces the
  action of extended Bargmann gravity constructed in
  \cite{Papageorgiou:2009zc,Bergshoeff:2016lwr,Hartong:2016yrf}.

\item[Example 3:]\ $(D,p;N_{0},N_{1})= (3,0;4,3)$. This example reproduces the extended Newtonian gravity algebra and the corresponding action constructed in
  \cite{Ozdemir:2019orp}.

\item[Example 4:]\ $(D,p;N_{0},N_{1})= (4,0;2,3)$. This is the example alluded
  to above where the first algebra $\mathfrak{g}(2,1)$ beyond the Galilei
  algebra cannot be used to construct an invariant action. Instead, in order
  to obtain an invariant action one has to proceed to the next order and
  bigger algebra $\mathfrak{g}(2,3)$. This is the same algebra that occurs in
  \cite{Hansen:2018ofj}.  We will compare the resulting action with the
    one of \cite{Hansen:2018ofj} in the conclusions.

  \item[Example 5:]$\ (D,p;N_{0},N_{1})= (4,1;2,1)$. This example reproduces
    the algebra and action of extended string Newton-Cartan gravity
    constructed in \cite{Bergshoeff:2018vfn}.

\end{description}

\begin{table}[!ht]
\renewcommand{\arraystretch}{1.5}
\begin{center}
\begin{tabular}{|c|c|c|c|}
\hline
\bf Algebra&$\mathbf{(D,p)=(3,0)}$& $\mathbf{(D,p)=(4,0)}$&$\mathbf{(D,p)=(4,1)}$ \\
\hline
&&&\\[-.6truecm]
$\mathfrak{g}(0,1)$&$\accentset{(0)}{B}$&$\accentset{(1)}{B}$&$\accentset{(0)}B$\\
$\mathfrak{g}(2,1)$&$\accentset{(2)}{B}$&$\accentset{(1)}{B}$&$\accentset{(2)}{B}$\\
$\mathfrak{g}(2,3)$&$\accentset{(2)}{B}$&$\accentset{(3)}{B}$&$\accentset{(2)}{B}$\\
$\mathfrak{g}(4,3)$&$\accentset{(4)}{B}$&$\accentset{(3)}{B}$&$\accentset{(4)}{B}$\\
\hline
\end{tabular}
\end{center}
\caption[]{{\small{In the table we list, for different values of $(D,p;N_{0},N_{1})$ the highest order term $\accentset{(n)}{B}$ that is invariant under the algebra $\mathfrak{g}(N_{0},N_{1})$ appearing in the first column. The first line corresponds to Galilei gravity that always
occurs as the lowest-order term in the Lie algebra expansion. In this
section we are going to discuss the five cases
$(D,p;N_{0},N_{1})=(3,0;2,1),\ (4,0;0,1),\ (4,0;2,3), (4,1;2,1)$ and
$(3,0;4,3)$.}}}\label{table:table2}
\end{table}

\subsection{4D Galilei Gravity}\label{sec:4DGalilei}

In this example we perform a Lie algebra expansion with a truncation to the
lowest-order Galilei algebra $\mathfrak{g}(0,1)$:\footnote{Note that this lowest-order truncation leads in  any dimension to a Galilei gravity theory.}
\begin{multicols}{2}
\begin{subequations}\label{expG}
\setlength{\abovedisplayskip}{-15pt}
\allowdisplaybreaks
\begin{align}
\tau_{\mu} &= \accentset{(0)}\tau_{\mu}\, \\
\Omega_{\mu}{}^{ab} &= \accentset{(0)}\Omega_{\mu}{}^{ab}\, \\
E_{\mu}{}^A &= \lambda\accentset{(1)}E_{\mu}{}^A\,, \\
\Omega_{\mu}{}^{0a} &= \lambda \accentset{(1)}\Omega_{\mu}{}^a\,.
\end{align}
\end{subequations}
\end{multicols}

This leads to the following curvatures:
\begin{subequations}
\begin{align}
  \accentset{(0)}{R}^{ab}(J)
  &
    = d\accentset{(0)}{\Omega}^{ab}-\accentset{(0)}{\Omega^{ac}}\wedge
    \accentset{(0)}{\Omega}^{b}_{\phantom{b}c}\, ,
  \\[4pt]
  \accentset{(0)}{R}^{0}(H)
  &
    = d\accentset{(0)}{\tau}\, ,
  \\[4pt]
  \accentset{(1)}{R}^{a}(G)
  &
    = d\accentset{(1)}{\Omega}^{a}
    +\accentset{(0)}{\Omega}^{ab}
    \wedge\accentset{(1)}{\Omega}_{b}\, ,
  \\[4pt]
  \accentset{(1)}{R}^{a}(P)
  &
    = d\accentset{(1)}{E}^{a}
    +\accentset{(0)}{\Omega}^{ab}\wedge \accentset{(1)}{E}_{b}
    -\accentset{(1)}{\Omega}^{a}\wedge \accentset{(0)}{\tau}\, .
\end{align}
\end{subequations}

Substituting the expansion \eqref{expG} into the Einstein-Hilbert action
\eqref{I1}, and taking the first-order term in the expansion, leads to the 4D
Galilei action
\begin{equation}\label{form1}
  S_{\rm Gal}
  = \int d^{4} x\, \epsilon^{\mu\nu\rho\sigma}\epsilon_{abc}
  \accentset{(1)}{E}_{\mu}{}^{a}\,\accentset{(0)}{\tau}_{\nu} \,
  \accentset{(0)}R_{\rho\sigma}{}^{bc}(J)\,.
\end{equation}
Note that the action does not contain the boost connection. This action is
invariant under local spatial rotations, Galilean boosts and general
coordinate transformations. Note that in this case the expansion of the
general coordinate transformations only leads to general coordinate
transformation but no further symmetries.

Using the identity
\begin{equation}
  \accentset{(3)}E
  =
  \textrm{det}\, (\accentset{(0)}{\tau}_{\mu},\accentset{(1)}E_{\mu}{}^{a})
  =
  \frac{1}{3!}\epsilon^{\mu\nu\rho\sigma}\epsilon_{abc}\,
  \accentset{(0)}{\tau}_{\mu} \accentset{(1)}{E}_{\nu}{}^{a}
  \accentset{(1)}E_{\rho}{}^{b}\accentset{(1)}E_{\sigma}{}^{c}
\end{equation}
this Galilei action can be rewritten in the alternative form\footnote{See
  eq. (4.24) of   \cite{Bergshoeff:2017btm}.}
\begin{equation}\label{form2}
  S_{\rm Gal}
  =
  \frac{1}{3!}\int d^{4} x\, \accentset{(3)}{E} \,\,
  \accentset{(-1)}{E}^{\mu}{}_{a}\, \accentset{(-1)}{E}^{\nu}{}_{b}\,
  \accentset{(0)}{R}_{\mu\nu}{}^{ab}(J)\,.
\end{equation}
where we have expanded the inverse Vierbein as
\begin{multicols}{2}
\begin{subequations}
\setlength{\abovedisplayskip}{-15pt}
\allowdisplaybreaks
\begin{align}
\tau^{\mu} &= \accentset{(0)}{\tau}^{\mu}\,, \\
E^{\mu}{}_{a} &= \lambda^{-1}\accentset{(-1)}{E}^{\mu}{}_{a}\,.
\end{align}
\end{subequations}
\end{multicols}

The equations of motion corresponding to the Galilei action, written in the
form \eqref{form1} or \eqref{form2}, are given by
\cite{Bergshoeff:2017btm}\footnote{Note that the last equation has only 12
  instead of 16 components as one naively would expect. The reason for this is
  that the 4D Galilei action is not only invariant under the 3 local Galilean
  boosts but also under 1 accidental local scale symmetry
  \cite{Bergshoeff:2017btm}.}
\begin{subequations}
\begin{eqnarray}
  \accentset{(-1)}{E}^{\mu}{}_{a}\ \accentset{(-1)}{E}^{\nu}{}_{b}
  \partial_{[\mu}\accentset{(0)}{\tau}_{\nu]}
  & = &
        0\,, \label{eq: RHab is zero}
  \\[.1truecm]
\tau^{\mu} \accentset{(-1)}{E}^{\nu}{}_{a}
  \partial_{[\mu}\accentset{(0)}{\tau}_{\nu]}
  & = &
        \frac{1}{2}\ \accentset{(-1)}R_{ab}{}^{b}(P)\,, \label{eq: Gal eom1}
  \\[.1truecm]
  \accentset{(-2)}{R}_{ab}{}^{c}(P)
  & = &
        - \delta_{[a}^{c} \accentset{(-2)}{R}_{b]d}{}^{d}(P)\,,\label{eq: Gal
        eom2}
  \\[.1truecm]
  \accentset{(-1)}{R}_{\mu b}{}^{ab}(J)
  & = &
        0 \,.
\end{eqnarray}
\end{subequations}

A notable feature of the Galilei action is that there is no second-order
formulation where all spin connections become dependent
\cite{Bergshoeff:2017btm}. One is left with an independent spin
connection component that acts as a Lagrange multiplier leading the
geometrical constraint that the geometry has twistless torsion, see
eq.~\eqref{eq: RHab is zero}.

\subsection{3D Extended Bargmann Gravity}\label{sec:3DextendedBargmann}

In this subsection we will show that the second example,
i.e.~$(D,p;N_{0},N_{1})= (3,0;2,1)$, leads to the action of 3D Extended
Bargmann gravity
constructed in  \cite{Papageorgiou:2009zc,Bergshoeff:2016lwr,Hartong:2016yrf}.\\

We write the 3-dimensional indices as $\hat{A}=\{0,a\}$, with
$a=1,2$, and  decompose the generators as follows:
\begin{subequations}
\begin{align}
J_{\hat{A}\hat{B}}&\rightarrow\{J, G_{a}\}\, ,\\
P_{\hat{A}}&\rightarrow\{P_{a}, H\}\, ,
\end{align}
\end{subequations}
where we have defined $G_{a}=J_{0a}$, $H=P_{0}$, and
$J=J_{12}=\tfrac{1}{2}\epsilon^{ab}J_{ab}$. The non-vanishing commutation
relations between these generators read
\begin{multicols}{2}
\begin{subequations}
\setlength{\abovedisplayskip}{-15pt}
\allowdisplaybreaks
\begin{align}
[G_{a},P_{b}] & = \delta_{ab}H\,, \\
[G_{a},H] & = P_{a}\, ,\\
[J,P_{a}] & = -\epsilon_{a}^{\ b}P_{b}\, ,\\
[J,G_{a}] & = -\epsilon_{a}^{\ b}G_{b}\, ,\\
[G_{a},G_{b}] & = -\epsilon_{ab}J\, .
\end{align}
\end{subequations}
\end{multicols}

We next introduce the following 1-form gauge fields associated to the generators
\begin{multicols}{2}
\begin{subequations}
\setlength{\abovedisplayskip}{-15pt}
\allowdisplaybreaks
\begin{align}
J&\rightarrow \Omega\, , \\
H&\rightarrow \tau\, ,\\
P_{a}&\rightarrow E^{a}\, ,\\
G_{a}&\rightarrow \Omega^{a}\, .
\end{align}
\end{subequations}
\end{multicols}

The corresponding curvatures are (omitting form indices)
\begin{subequations}
\begin{align}
R(J) & = d\Omega-\epsilon_{ab}\Omega^{a}\wedge\Omega^{b}\, ,\\
R(H) & = d\tau+\delta_{ab}\Omega^{a}\wedge E^{b}\, ,\\[4pt]
R^{a}(P) & = dE^{a}+\Omega^{a}\wedge\tau-\epsilon^{a}_{\ b}\Omega\wedge E^{b}\, ,\\
R^{a}(G) & = d\Omega^{a}-\epsilon^{a}_{\ b}\Omega\wedge\Omega^{b}\, .
\end{align}
\end{subequations}

We now expand the Lie algebra with respect to the decomposition
$\mathfrak{g}=V_{0}\oplus V_{1}$ with
\begin{equation}\label{V0}
V_{0}=\{J,H\} ,\,\,\,\, \text{and}\,\,\,\,
V_{1}=\{P_{a},G_{a}\}\, .
\end{equation}

Imposing the consistent truncation $(N_{0},N_{1}) = (2,1)$, the expansions of
the gauge fields are given by
\begin{multicols}{2}
\begin{subequations}
\setlength{\abovedisplayskip}{-15pt}
\allowdisplaybreaks
\begin{align}
\Omega & = \accentset{(0)}{\Omega}+\lambda^{2} \accentset{(2)}{\Omega} \, ,\\
\tau & = \accentset{(0)}{\tau}+\lambda^{2} \accentset{(2)}{\tau} \, ,\\
E^{a} & = \lambda\accentset{(1)}{E}^{a} \, , \\
\Omega^{a} & = \lambda\accentset{(1)}{\Omega}^{a} \, .
\end{align}
\end{subequations}
\end{multicols}

The corresponding curvature 2-forms are given by:
\begin{multicols}{2}
\begin{subequations}
\setlength{\abovedisplayskip}{-15pt}
\allowdisplaybreaks
\begin{align}
  \accentset{(0)}{R}(J)
  &=
    d\accentset{(0)}{\Omega}\, ,
  \\
  \accentset{(0)}{R}(H)
  &=
    d\accentset{(0)}{\tau}\, ,
  \\
  \accentset{(2)}{R}(J)
  &=
    d\accentset{(2)}{\Omega}
    -\epsilon_{ab}\accentset{(1)}{\Omega}^{a}\wedge\accentset{(1)}{\Omega}^{b}\,,
  \\
  \accentset{(2)}{R}(H)
  &=
    d\accentset{(2)}{\tau}
    +\delta_{ab}\accentset{(1)}{\Omega}^{a}\wedge \accentset{(1)}{E}^{b}\,,
  \\
  \accentset{(1)}{R}^{a}(G)
  &=
    d\accentset{(1)}{\Omega}^{a}
    -\epsilon^{a}_{\ b}\accentset{(0)}{\Omega}\wedge
    \accentset{(1)}{\Omega}^{b}\,  ,
  \\
  \accentset{(1)}{R}^{a}(P)
  &=
    d\accentset{(1)}{E}^{a}+\accentset{(1)}{\Omega}^{a}\wedge
    \accentset{(0)}{\tau}
    -\epsilon^{a}_{\ b}\accentset{(0)}{\Omega}\wedge \accentset{(1)}{E}^{b}\, .
\end{align}
\end{subequations}
\end{multicols}

In order to construct an action, we consider the 3D Einstein-Hilbert
Lagrangian density given by
\begin{align}
  B
  &
    = \epsilon_{\hat{A}\hat{B}\hat{C}}R^{\hat{A}\hat{B}}(J)\wedge E^{\hat{C}}
   =
    2R(J)\wedge\tau+2\epsilon_{ab}R^{a}(G)\wedge E^{b}\, .\label{eq:EH3Daction}
\end{align}

Expanding order by order we obtain for the lowest-order term
\begin{equation}
   \accentset{(0)}{B}
    =
    \accentset{(0)}{R}(J)\wedge\accentset{(0)}{\tau}\, .
\end{equation}

This is precisely the action of 3D Galilei gravity, see eq.~(4.44) of
\cite{Bergshoeff:2017btm}.  The underlying algebra $\mathfrak{g}(0,1)$
corresponding to this truncation is the Galilei algebra which is the
In\"on\"u-Wigner contraction of the Poincar\'e algebra with respect to the
subalgebra $V_{0}$ defined in eq.~\eqref{V0}.  This is the 3D analogue of the
first example.

Next, we consider the second-order term
\begin{align}
    \accentset{(2)}{B}
    =
    2\accentset{(0)}{R}(J) \wedge \accentset{(2)}{\tau}
    +2\accentset{(2)}{R}(J)\wedge \accentset{(0)}{\tau}
    +2 \epsilon_{ab}\accentset{(1)}{R}^{a}(G)\wedge \accentset{(1)}{E}^{b}\, .
\end{align}

Relabeling the gauge fields  as
\begin{multicols}{3}
\begin{subequations}
\setlength{\abovedisplayskip}{-15pt}
\allowdisplaybreaks
\begin{align}
\accentset{(0)}{\Omega} & = \Omega\, ,\\
\accentset{(2)}{\Omega}  &= s\, ,\\
\accentset{(0)}{\tau}  &= \tau\, , \\
\accentset{(2)}{\tau}  &= m\, , \\
\accentset{(1)}{\Omega}^{a}  &= \Omega^{a}\, , \\
\accentset{(1)}{E}^{a}  &= E^{a}\, ,
\end{align}
\end{subequations}
\end{multicols}
\noindent
this term reads
\begin{equation}
    \accentset{(2)}{B}
    =
    2R(J)\wedge m+2R(S)\wedge \tau+2  \epsilon_{ab}R^{a}(G)\wedge E^{b}\, ,
\end{equation}
which is precisely the Lagrangian of 3D extended Bargmann gravity
\cite{Papageorgiou:2009zc,Bergshoeff:2016lwr,Hartong:2016yrf}.

The corresponding algebra $\mathfrak{g}(2,1)$ that occurs in the Lie algebra
expansion is given by
\begin{multicols}{2}
\begin{subequations}
\setlength{\abovedisplayskip}{-15pt}
\allowdisplaybreaks
\begin{align}
  [\accentset{(1)}{G}_{a},\accentset{(1)}{P}_{b}]
  & = \delta_{ab}\accentset{(2)}{H}\, ,
  \\
  [\accentset{(1)}{G}_{a},\accentset{(0)}{H}]
  & = \accentset{(1)}{P}_{a}\, ,
  \\
  [\accentset{(0)}{J},\accentset{(1)}{P}_{a}]
  & =
    -\epsilon_{ab}\accentset{(1)}{P}^{a}\, ,
  \\
  [\accentset{(0)}{J},\accentset{(1)}{G}_{a}]
  & =
    -\epsilon_{ab}\accentset{(1)}{G}^{a}\, ,
  \\
  [\accentset{(1)}{G}_{a},\accentset{(1)}{G}_{b}]
  & =
    -\epsilon_{ab}\accentset{(2)}{J}\, .
\end{align}
\end{subequations}
\end{multicols}

Identifying $\accentset{(2)}{J}=S$ and $\accentset{(2)}{H}=Z$ and using for
the lowest order terms the labels of the initial field, the non-zero
commutators of the algebra read
\begin{multicols}{2}
\begin{subequations}
\setlength{\abovedisplayskip}{-15pt}
\allowdisplaybreaks
\begin{align}
[G_{a},P_{b}] & = \delta_{ab}Z\, , \\
[G_{a},H] & = P_{a}\, , \\
[J,P_{a}] & = -\epsilon_{ab}P^{a}\, , \\
[J,G_{a}] & = -\epsilon_{ab}G^{a}\, , \\
[G_{a},G_{b}] & = -\epsilon_{ab}S\, .
\end{align}
\end{subequations}
\end{multicols}
\noindent
which is the extended Bargmann algebra of
\cite{Papageorgiou:2009zc,Bergshoeff:2016lwr,Hartong:2016yrf}. One may verify
that the transformation rules obtained by our procedure coincide with the ones
of \cite{Papageorgiou:2009zc,Bergshoeff:2016lwr,Hartong:2016yrf}.

Note that in this example the expansion of the general coordinate
transformations leads to further symmetries but these are related, up to
trivial symmetries, to the expanded $P$-transformations. For instance,
expanding
\begin{equation}
\xi^{\mu} = \accentset{(0)}{\xi}^{\mu} + \lambda^{2}\accentset{(2)}{\xi}^{\mu}
\end{equation}
one finds that the gauge field $\accentset{(2)}{\tau}_{\mu}$, up to a trivial
symmetry of the expanded action, transforms as
$\delta \accentset{(2)}{\tau}_{\mu} =
\partial_{\mu}\big(\accentset{(2)}{\xi}^{\lambda}\accentset{(0)}{\tau}_{\lambda})$
which is precisely an expanded $P$ transformation with
$\accentset{(2)}\eta^{\,0} =
\accentset{(2)}{\xi}^{\lambda}\accentset{(0)}{\tau}_{\lambda}$.
\subsection{3D Extended Newtonian Gravity}

In this subsection we study the case $(D,p;N_{0},N_{1})= (3,0;4,3)$.  We show
that this corresponds to the algebra given  in \cite{Ozdemir:2019orp} and we
reproduce the corresponding action.

We consider the same setting as in
subsection~\ref{sec:3DextendedBargmann} with the same subspace decomposition
of the algebra, but now we truncate the expansion of the gauge fields at order
$(N_{0},N_{1})=(4,3)$. Explicitly, we have
\begin{multicols}{2}
\begin{subequations}
\setlength{\abovedisplayskip}{-15pt}
\allowdisplaybreaks
\begin{align}
\Omega & = \accentset{(0)}{\Omega}+\lambda^{2} \accentset{(2)}{\Omega}+\lambda^{4} \accentset{(4)}{\Omega} \, ,\\
\tau & = \accentset{(0)}{\tau}+\lambda^{2} \accentset{(2)}{\tau}+\lambda^{4} \accentset{(4)}{\tau} \, ,\\
E^{a} & = \lambda\accentset{(1)}{E}^{a}+ \lambda^{3}\accentset{(3)}{E}^{a}\, , \\
\Omega^{a} & = \lambda\accentset{(1)}{\Omega}^{a}+\lambda^{3}\accentset{(3)}{\Omega}^{a} \, .
\end{align}
\end{subequations}
\end{multicols}

The corresponding curvature 2-forms are given by:
\begin{subequations}
\begin{multicols}{2}
\setlength{\abovedisplayskip}{-15pt}
\allowdisplaybreaks
\begin{align}
  \accentset{(0)}{R}(J)
  & = d\accentset{(0)}{\Omega}\, ,
  \\
  \accentset{(0)}{R}(H)
  & = d\accentset{(0)}{\tau}\, ,
  \\
  \accentset{(2)}{R}(J)
  & =
    d\accentset{(2)}{\Omega}-\epsilon_{ab}\accentset{(1)}{\Omega}^{a}\wedge\accentset{(1)}{\Omega}^{b}\,,
  \\
  \accentset{(2)}{R}(H)
  & =
    d\accentset{(2)}{\tau}+\delta_{ab}\accentset{(1)}{\Omega}^{a}\wedge
    \accentset{(1)}{E}^{b}\,,
  \\
  \accentset{(4)}{R}(J)
  & =
    d\accentset{(4)}{\Omega}-2\epsilon_{ab}\accentset{(1)}{\Omega}^{a}\wedge\accentset{(3)}{\Omega}^{b}\,,
  \\
  \accentset{(4)}{R}(H)
  & =
    d\accentset{(2)}{\tau}+\delta_{ab}\accentset{(1)}{\Omega}^{a}\wedge
    \accentset{(3)}{E}^{b}+\delta_{ab}\accentset{(3)}{\Omega}^{a}\wedge
    \accentset{(1)}{E}^{b}\,,
  \\
  \accentset{(1)}{R}^{a}(G)
  & =
    d\accentset{(1)}{\Omega}^{a}-\epsilon^{a}_{\
    b}\accentset{(0)}{\Omega}\wedge\accentset{(1)}{\Omega}^{b}\,  ,
  \\
  \accentset{(1)}{R}^{a}(P)
  & =
    d\accentset{(1)}{E}^{a}+\accentset{(1)}{\Omega}^{a}\wedge\accentset{(0)}{\tau}-\epsilon^{a}_{\
    b}\accentset{(0)}{\Omega}\wedge \accentset{(1)}{E}^{b}\, ,
  \\
  \accentset{(3)}{R}^{a}(G)
  & = d\accentset{(3)}{\Omega}^{a}-\epsilon^{a}_{\
    b}\accentset{(0)}{\Omega}\wedge\accentset{(3)}{\Omega}^{b}-\epsilon^{a}_{\
    b}\accentset{(2)}{\Omega}\wedge\accentset{(1)}{\Omega}^{b}\,  ,
  \\
  \accentset{(3)}{R}^{a}(P)
  & =
    d\accentset{(3)}{E}^{a}+\accentset{(3)}{\Omega}^{a}\wedge\accentset{(0)}{\tau}+\accentset{(1)}{\Omega}^{a}\wedge\accentset{(2)}{\tau}
    \nonumber\\
  &
    -\epsilon^{a}_{\ b}\accentset{(0)}{\Omega}\wedge
    \accentset{(3)}{E}^{b}-\epsilon^{a}_{\ b}\accentset{(2)}{\Omega}\wedge
    \accentset{(1)}{E}^{b\, }.
\end{align}
\end{multicols}
\end{subequations}

We consider the Einstein-Hilbert action eq.~\eqref{eq:EH3Daction} and consider
the expansion defined above, but now we focus on the fourth-order term
$\accentset{(4)}{B}$ which for the truncation we consider defines an invariant action.
We obtain
\begin{align}
  \accentset{(4)}{B}
  =
  \epsilon_{ab}\left[\accentset{(1)}{R}^{a}(G)\wedge \accentset{(3)}{E}^{b}
  +\accentset{(3)}{R}^{a}(G)\wedge \accentset{(1)}{E}^{b}
  +\accentset{(0)}{R}^{ab}(J)\wedge \accentset{(4)}{\tau}
  +\accentset{(2)}{R}^{ab}(J)\wedge \accentset{(2)}{\tau}
  +\accentset{(4)}{R}^{ab}(J)\wedge \accentset{(0)}{\tau}\right]\, .
\end{align}
Relabeling the gauge fields  as
\begin{multicols}{3}
\begin{subequations}
\setlength{\abovedisplayskip}{-15pt}
\allowdisplaybreaks
\begin{align}
\accentset{(0)}{\Omega} & = \omega\, ,\\
\accentset{(2)}{\Omega}   & =  s\, ,\\
\accentset{(4)}{\Omega}   & =  z\, ,\\
\accentset{(0)}{\tau}   & =  \tau\, , \\
\accentset{(2)}{\tau}   & =  m\, , \\
\accentset{(4)}{\tau}   & =  -y\, , \\
\accentset{(1)}{\Omega}^{a}   & =  \omega^{a}\, , \\
\accentset{(3)}{\Omega}^{a}   & =  b^{a}\, , \\
\accentset{(1)}{E}^{a}   & =  \epsilon^{ab}e_{b}\, ,\\
\accentset{(3)}{E}^{a}   & =  \epsilon^{ab}t_{b},
\end{align}
\end{subequations}
\end{multicols}
\noindent
the action above takes the form
\begin{align}
  \accentset{(4)}{B}=R(S)\wedge m +R(Z)\wedge \tau-R(J)\wedge y
  -R^{a}(G)\wedge t_{a}-R^{a}(B)\wedge e_{a},
\end{align}
where we have used the notation $R(J)=\epsilon_{ab}R^{ab}(J)$ and the same for
the curvatures associated with $S$ and $Z$. This action corresponds to the
action appearing in \cite{Ozdemir:2019orp}.

Let us take a closer look to the algebra $\mathfrak{g}(4,3)$. The non-zero
commutators of the algebra $\mathfrak{g}(4,3)$ are given explicitly by
\begin{multicols}{2}
\begin{subequations}
\setlength{\abovedisplayskip}{-15pt}
\allowdisplaybreaks
\begin{align}
[\accentset{(0)}{J}_{ab},\accentset{(1)}{G}_{d}] & = 2\eta_{d[b}\accentset{(1)}{G}_{a]}\\
[\accentset{(0)}{J}_{ab},\accentset{(3)}{G}_{d}] & = 2\eta_{d[b}\accentset{(3)}{G}_{a]}\\
[\accentset{(2)}{J}_{ab},\accentset{(1)}{G}_{d}] & = 2\eta_{d[b}\accentset{(3)}{G}_{a]}\\
[\accentset{(1)}{G}_{a},\accentset{(1)}{G}_{b}] & = \accentset{(2)}{J}_{ab}\\
[\accentset{(1)}{G}_{a},\accentset{(3)}{G}_{b}] & = \accentset{(4)}{J}_{ab}\\
[\accentset{(0)}{J}_{ab},\accentset{(1)}{P}_{d}] & = 2\eta_{d[b}\accentset{(1)}{P}_{a]}\\
[\accentset{(0)}{J}_{ab},\accentset{(3)}{P}_{d}] & = 2\eta_{d[b}\accentset{(3)}{P}_{a]}\\
[\accentset{(2)}{J}_{ab},\accentset{(1)}{P}_{d}] & = 2\eta_{d[b}\accentset{(3)}{P}_{a]}\\
[\accentset{(1)}{G}_{a},\accentset{(0)}{H}] & = \accentset{(1)}{P}_{a}\\
[\accentset{(1)}{G}_{a},\accentset{(2)}{H}] & = \accentset{(3)}{P}_{a}\\
[\accentset{(3)}{G}_{a},\accentset{(0)}{H}] & = \accentset{(3)}{P}_{a}\\
[\accentset{(1)}{G}_{a},\accentset{(1)}{P}_{b}] & = \eta_{ab}\accentset{(2)}{H}\\
[\accentset{(1)}{G}_{a},\accentset{(3)}{P}_{b}] & = \eta_{ab}\accentset{(4)}{H}\\
[\accentset{(3)}{G}_{a},\accentset{(1)}{P}_{b}] & = \eta_{ab}\accentset{(4)}{H}.
\end{align}
\end{subequations}
\end{multicols}

With the identifications
\begin{multicols}{3}
\begin{subequations}
\setlength{\abovedisplayskip}{-15pt}
\allowdisplaybreaks
\begin{align}
\accentset{(0)}{J}_{ab} & = \epsilon_{ab}J,\\
\accentset{(2)}{J}_{ab} & = \epsilon_{ab}S,\\
\accentset{(4)}{J}_{ab} & = \epsilon_{ab}Z,\\
\accentset{(0)}{H} & = H,\\
\accentset{(2)}{H} & = M,\\
\accentset{(4)}{H} & = -Y,\\
\accentset{(1)}{P}_{a} & = \epsilon_{ab}P^{b},\\
\accentset{(3)}{P}_{a} & = \epsilon_{ab}T^{b},\\
\accentset{(1)}{G}_{a} & = G_{a},\\
\accentset{(3)}{G}_{a} & = B_{a},
\end{align}
\end{subequations}
\end{multicols}
\noindent
the algebra above reproduces exactly the algebra of \cite{Ozdemir:2019orp}:
\begin{multicols}{2}
\begin{subequations}
\setlength{\abovedisplayskip}{-15pt}
\allowdisplaybreaks
\begin{align}
[J,G_{a}] & = -\epsilon_{ab}G^{b},\\
[J,B_{a}] & = -\epsilon_{ab}B^{b},\\
[S,G_{a}] & = -\epsilon_{ab}B^{b},\\
[G_{a},G_{b}] & = \epsilon_{ab}S,\\
[G_{a},B_{b}] & = \epsilon_{ab}Z,\\
[J,P_{a}] & = -\epsilon_{ab}P^{b},\\
[J,T_{a}] & = -\epsilon_{ab}T^{b},\\
[S,P_{a}] & = -\epsilon_{ab}T^{b},\\
[G_{a},H] & = \epsilon_{ab}P^{b},\\
[G_{a},M] & = \epsilon_{ab}T^{b},\\
[B_{a},H] & = \epsilon_{ab}T^{b},\\
[G_{a},P_{b}] & = \epsilon_{ab}M,\\
[G_{a},T_{b}] & = -\epsilon_{ab}Y,\\
[B_{a},P_{b}] & = -\epsilon_{ab}Y.
\end{align}
\end{subequations}
\end{multicols}

\subsection{Beyond 4D Galilei Gravity}

In this subsection we consider the example $(D,p;N_{0},N_{1})= (4,0;2,3)$.
Since the indices $A,B,C\ldots$ only take the value $0$, we will suppress
them. The decomposition of the generators in this case read
\begin{subequations}
\begin{align}
J_{\hat{A}\hat{B}}&\rightarrow\{J_{ab}, G_{a}\}\, , \\
P_{\hat{A}}&\rightarrow\{H, P_{a}\}\, ,
\end{align}
\end{subequations}
with $a=1,2,3$. The non-vanishing commutators between these generators are given by
\begin{multicols}{2}
\begin{subequations}
\setlength{\abovedisplayskip}{-15pt}
\allowdisplaybreaks
\begin{align}
[G_{a},P_{b}] & = \delta_{ab}H\, , \\
[G_{a},H] & = P_{a}\, , \\
[J_{ab},P_{c}] & = \delta_{bc}P_{a}-\delta_{ac}P_{b}\, , \\
[J_{ab},G_{c}] & = \delta_{bc}G_{a}-\delta_{ac}G_{b}\, , \\
[J_{ab},J_{cd}] & = 4\delta_{[a[c}J_{d]b]}\, , \\
[G_{a},G_{b}] & = J_{ab}\, .
\end{align}
\end{subequations}
\end{multicols}

We associate the following gauge fields to these curvatures:
\begin{multicols}{2}
\begin{subequations}
\setlength{\abovedisplayskip}{-15pt}
\allowdisplaybreaks
\begin{align}
J_{ab}&\rightarrow \Omega^{ab}\, , \\
P_{a}&\rightarrow E^{a}\, , \\
G_{a}&\rightarrow \Omega^{a}\, , \\
H&\rightarrow \tau\,.
\end{align}
\end{subequations}
\end{multicols}

The corresponding  curvatures are defined by
\begin{subequations}
\begin{align}
R^{ab}(J) & = d\Omega^{ab}-\Omega^{ac}\wedge \Omega^{b}_{\phantom{b}c}+\Omega^{a}\wedge\Omega^{b}\, , \\
R^{a}(G) & = d\Omega^{a}+\Omega^{ab}\wedge\Omega_{b}\, , \\
R^{a}(P) & = dE^{a}+\Omega^{ab}\wedge E_{b}+\Omega^{a}\wedge \tau\, , \\
R^{A}(H) & = d\tau+\Omega^{a}\wedge E_{a}\, .
\end{align}
\end{subequations}

We now consider a Lie algebra expansion according to the  decomposition
$\mathfrak{g}=V_{0}\oplus V_{1}$ with
\begin{equation}
V_{0} = \{J_{ab},H\}\, , \,\,\,\, \text{and}\,\,\,\,
V_{1}  = \{G_{a},P_{a}\}\, ,
\end{equation}

Following Table~\ref{table:table2}, we impose the truncation
$(N_{0},N_{1})=(2,3)$ in order to construct the first action that goes beyond
the lowest-order Galilei gravity action discussed in the first example.  This
particular truncation induces the following expansion
\begin{multicols}{2}
\begin{subequations}
\setlength{\abovedisplayskip}{-15pt}
\begin{align}
\Omega^{ab} & = \accentset{(0)}{\Omega}^{ab}+\lambda^{2}\accentset{(2)}{\Omega}^{ab}\ , \\
\tau & = \accentset{(0)}{\tau}+\lambda^{2}\accentset{(2)}{\tau}\ , \\
\Omega^{a} & = \lambda\accentset{(1)}{\Omega}^{a}+\lambda^{3}\accentset{(3)}{\Omega}^{a}\ , \\
E^{a} & = \lambda\accentset{(1)}{E}^{a}+\lambda^{3}\accentset{(3)}{E}^{a}\ ,
\end{align}
\end{subequations}
\end{multicols}

We now substitute this expansion into the first-order Einstein-Hilbert
Lagrangian density 4-form
\begin{align}
  B
  & =
    \epsilon_{\hat{A}\hat{B}\hat{C}\hat{D}}R^{\hat{A}\hat{B}}\wedge
    E^{\hat{C}}\wedge E^{\hat{D}}
    =2\epsilon_{abc}\left(R^{a}(G)\wedge E^{b}\wedge E^{c}-R^{ab}(J)\wedge
    E^{c}\wedge \tau\right)\, ,
\end{align}
adn we find at lowest order
\begin{equation}
  \accentset{(1)}{B}
    =
    -2\epsilon_{abc}
    \accentset{(0)}{R}^{ab}(J)\wedge\accentset{(1)}{E}^{c}\wedge \accentset{(0)}{\tau}\, .
\end{equation}

As expected, this corresponds precisely to the action \eqref{form1} of 4D
Galilei gravity discussed in the first example.

It is only when we consider the truncation $(N_{0},N_{1})=(2,3)$, but not the
stronger truncation $(N_{0},N_{1})=(2,1)$. that the next order term
$\accentset{(3)}B$ also leads to an invariant action:

\begin{eqnarray}\label{beyond}
 \accentset{(3)}{B}
     & = &  2\epsilon_{abc}
    \accentset{(1)}{R}^{a}(G)\wedge \accentset{(1)}{E}^{b}\wedge\accentset{(1)}{E}^{c}
    -2\epsilon_{abc}
    \accentset{(2)}{R}^{ab}(J)\wedge\accentset{(1)}{E}^{c}\wedge \accentset{(0)}{\tau}\nonumber\\[6pt]
   &&\hskip 0truecm  -2\epsilon_{abc}
    \accentset{(0)}{R}^{ab}(J)\wedge\accentset{(1)}{E}^{c}\wedge\accentset{(2)}{\tau}
    -2\epsilon_{abc}
    \accentset{(0)}{R}^{ab}(J)\wedge\accentset{(3)}{E}^{c}\wedge\accentset{(0)}{\tau}\,.
    \end{eqnarray}
\vskip .2truecm

\noindent
This is the first action in the Lie algebra expansion that goes beyond 4D
Galilei gravity. Note that, although this action does not depend on
$\accentset{(3)}{\Omega}^{a}$, it is invariant under the gauge transformations
of parameter $\accentset{(3)}{\lambda}^{a}$ (see eq
\eqref{eq:curvaturesB}). This is similar to the situation described in Section
\ref{sec:4DGalilei}.

The definition of the different curvatures and the transformation rules of the
gauge fields and curvatures follow from the Lie algebra expansion. For the
convenience of the reader, we have given the explicit expressions in
Appendix~\ref{appendix}.

The algebra $\mathfrak{g}(2,3)$ underlying the action \eqref{beyond} is given by
\begin{multicols}{2}
\begin{subequations}
\setlength{\abovedisplayskip}{-15pt}
\begin{align}
[\accentset{(1)}{G}_{a},\accentset{(1)}{P}_{b}] & = \delta_{ab}\accentset{(2)}{H}\\
[\accentset{(1)}{G}_{a},\accentset{(0)}{H}] & = \accentset{(1)}{P}_{a}\\
[\accentset{(3)}{G}_{a},\accentset{(0)}{H}] & = \accentset{(3)}{P}_{a}\\
[\accentset{(1)}{G}_{a},\accentset{(2)}{H}] & = \accentset{(3)}{P}_{a}\\
[\accentset{(0)}{J}_{ab},\accentset{(1)}{P}_{c}] & = \delta_{bc}\accentset{(1)}{P}_{a}-\delta_{ac}\accentset{(1)}{P}_{b}\\
[\accentset{(0)}{J}_{ab},\accentset{(3)}{P}_{c}] & = \delta_{bc}\accentset{(3)}{P}_{a}-\delta_{ac}\accentset{(3)}{P}_{b}\\
[\accentset{(2)}{J}_{ab},\accentset{(1)}{P}_{c}] & = \delta_{bc}\accentset{(3)}{P}_{a}-\delta_{ac}\accentset{(3)}{P}_{b}\\
[\accentset{(0)}{J}_{ab},\accentset{(1)}{G}_{c}] & = \delta_{bc}\accentset{(1)}{G}_{a}-\delta_{ac}\accentset{(1)}{G}_{b}\\
[\accentset{(0)}{J}_{ab},\accentset{(3)}{P}_{c}] & = \delta_{bc}\accentset{(3)}{G}_{a}-\delta_{ac}\accentset{(3)}{G}_{b}\\
[\accentset{(2)}{J}_{ab},\accentset{(1)}{G}_{c}] & =  \delta_{bc}\accentset{(3)}{G}_{a}-\delta_{ac}\accentset{(3)}{G}_{b}\\
[\accentset{(0)}{J}_{ab},\accentset{(0)}{J}_{cd}]& = 4\delta_{[a[c}\accentset{(0)}{J}_{d]b]}\\
[\accentset{(0)}{J}_{ab},\accentset{(2)}{J}_{cd}] & = 4\delta_{[a[c}\accentset{(2)}{J}_{d]b]}\\
[\accentset{(1)}{G}_{a},\accentset{(1)}{G}_{b}] & = \accentset{(2)}{J}_{ab}
\end{align}
\end{subequations}
\end{multicols}

Upon making the identifications
\begin{multicols}{4}
\begin{subequations}
\setlength{\abovedisplayskip}{-15pt}
\allowdisplaybreaks
\begin{align}
  N=&\accentset{(2)}{H}\, ,\\
  S_{ab}=&\accentset{(2)}{J}_{ab}\, , \\
  T_{a}=&\accentset{(3)}{P}_{a}\, , \\
  B_{a}=&\accentset{(3)}{G}_{a}\, ,
\end{align}
\end{subequations}
\end{multicols}

\noindent
and assigning the lowest-order terms the same labels as the original
generators, this is precisely the algebra appearing in
\cite{Hansen:2018ofj}. Note that we are using here a first-order formulation
of the action as opposed to the second-order formulation used in
\cite{Hansen:2018ofj}. We will discuss and compare the two actions in the
Conclusions.

\subsection{4D Extended String Newton-Cartan Gravity}

The last example we consider corresponds to the case
$(D,p;N_{0},N_{1})= (4,1;2,1)$.  It will lead to an extension of the String Newton-Cartan Gravity theory that underlies nonrelativistic string theory. In this case we decompose the generators of
the Poincar\'e algebra as follows:
\begin{subequations}
  \begin{align}
    J_{\hat{A}\hat{B}}& \rightarrow\{M, G_{Aa},J\}\, ,\\
    P_{\hat{A}} &\rightarrow\{H_{A}, P_{a}\}\, ,
  \end{align}
\end{subequations}

where $A=\{0,1\}$ and $a=\{2,3\}$.\footnote{We use the conventions
  $\epsilon_{23}=+1,\epsilon_{01}=+1$ for $\epsilon_{ab}$ and
  $\epsilon_{AB}$. Furthermore, we have defined
  $J=J^{23}=\tfrac{1}{2}\epsilon_{ab}J^{ab}$ and
  $M=J^{01}=\tfrac{1}{2}\epsilon_{AB}J^{AB}$.} The non-vanishing commutation
relations in terms of these generators are given by
\begin{multicols}{2}
\begin{subequations}
\setlength{\abovedisplayskip}{-15pt}
\allowdisplaybreaks
\begin{align}
[M,H_{A}] & = -\epsilon_{A}^{\ B}H_{B}\, ,\\
[J,P_{a}] & = -\epsilon_{a}^{\ b}P_{b}\, , \\
[H_{A},G_{ba}] & = \eta_{AB}P_{a}\, ,\\
[M,G_{Aa}] & = -\epsilon_{A}^{\ B}G_{Ba}\, , \\
[J,G_{Aa}] & = -\epsilon_{a}^{\ b}G_{Ab}\, ,\\
[G_{Aa},P_{b}] & = \delta_{ab}H_{A}\, , \\
[G_{Aa},G_{Bb}] & = -\eta_{AB}\epsilon_{ab}J+\eta_{ab}\epsilon_{AB}M\, .
\end{align}
\end{subequations}
\end{multicols}

The corresponding gauge fields are renamed as follows:
\begin{multicols}{3}
\begin{subequations}
\setlength{\abovedisplayskip}{-15pt}
\allowdisplaybreaks
\begin{align}
J&\rightarrow\ \Omega\, ,\\
M&\rightarrow\ \Sigma\, , \\
G_{Ab}&\rightarrow\ \Omega^{Ab}\, , \\
P_{a}&\rightarrow\ E^{a}\, , \\
H_{A}&\rightarrow\ \tau^{A}\, .
\end{align}
\end{subequations}
\end{multicols}

The  expanded curvatures of these gauge fields are defined by
\begin{subequations}
\begin{align}
  R(J)
  &
    = d\Omega-\eta_{AB}\epsilon_{ab}\Omega^{Aa}\wedge\Omega^{Bb}\, ,
  \\[4pt]
  R(M)
  &
    = d\Sigma+\eta_{ab}\epsilon_{AB}\Omega^{Aa}\wedge\Omega^{Bb}\, ,
  \\[4pt]
  R^{A}(H)
  &
    = d\tau^{A}-\epsilon_{AB}\Sigma\wedge
    \tau^{B}+\delta_{ab}\Omega^{Aa}\wedge E^{b}\, ,
  \\[4pt]
  R^{a}(P)
  &
    = dE^{a}-\epsilon^{ab}\Omega \wedge E_{b}-\eta_{AB}\Omega^{Aa}\wedge
    \tau^{B}\, ,
  \\[4pt]
  R^{Aa}(G)
  &
    = d\Omega^{Aa}-\epsilon^{ab}\Omega\wedge\Omega^{A}_{\ b}
    +\epsilon^{AB}\Omega\wedge\Omega^{\ a}_{B}\, .
\end{align}
\end{subequations}

We now consider the Lie algebra expansion with respect to the following
decomposition $\mathfrak{g}=V_{0}\oplus V_{1}$ with
\begin{equation}
V_{0}=\{J, M, H_{A}\} ,\,\,\,\,\, \text{and}\,\,\,\,
V_{1}=\{P_{a}, G_{Aa}\}\, ,
\end{equation}
The truncation $(N_{0},N_{1})=(2,1)$  induces the finite  expansions
\begin{multicols}{2}
\begin{subequations}
\setlength{\abovedisplayskip}{-15pt}
\allowdisplaybreaks
\begin{align}
  \Omega & =  \accentset{(0)}{\Omega}+\lambda^{2}\ \accentset{(2)}{\Omega}\, , \\
  \Sigma & =  \accentset{(0)}{\Sigma}+\lambda^{2}\ \accentset{(2)}{\Sigma}\, , \\
  \tau^{A} & =  \accentset{(0)}{\tau}^{A}+\lambda^{2}\ \accentset{(2)}{\tau}^{A}\, , \\
   \Omega^{Ab}& = \lambda\ \accentset{(1)}{\Omega}^{Ab}\, ,\\
  E^{a} & =  \lambda\ \accentset{(1)}{E}^{a}\, .
\end{align}
\end{subequations}
\end{multicols}

The curvatures of these gauge fields are given by
\begin{subequations}
\begin{align}
  \accentset{(0)}{R}(J) & =  d\accentset{(0)}{\Omega}\, ,\\[4pt]
  \accentset{(2)}{R}(J) & =  d\accentset{(2)}{\Omega}
    -\eta_{AB}\epsilon_{ab}\accentset{(1)}{\Omega}^{Aa}
    \wedge\accentset{(1)}{\Omega}^{Bb}\, ,
  \\[4pt]
  \accentset{(0)}{R}(M)& = d\accentset{(0)}{\Sigma}\, ,\\[4pt]
  \accentset{(2)}{R}(M)& =
    d\accentset{(2)}{\Sigma}+\eta_{ab}\epsilon_{AB}\accentset{(1)}{\Omega}^{Aa}
    \wedge\accentset{(1)}{\Omega}^{Bb}\, ,
  \\[4pt]
  \accentset{(0)}{R}^{A}(H)
  &
    = d\accentset{(0)}{\tau}^{A}
    -\epsilon_{AB}\accentset{(0)}{\Sigma}\wedge \accentset{(0)}{\tau}^{B}\, ,
  \\[4pt]
  \accentset{(2)}{R}^{A}(H)
  &
    = d\accentset{(2)}{\tau}^{A}
    -\epsilon_{AB}\accentset{(2)}{\Sigma}\wedge \accentset{(0)}{\tau}^{B}
    -\epsilon_{AB}\accentset{(0)}{\Sigma}\wedge \accentset{(2)}{\tau}^{B}
    +\delta_{ab}\accentset{(1)}{\Omega}^{Aa}\wedge \accentset{(1)}{E}^{b}\, ,
  \\
  \accentset{(1)}{R}^{a}(P)
  &
    = d\accentset{(1)}{E}^{a}
    -\epsilon^{ab}\accentset{(0)}{\Omega} \wedge \accentset{(1)}{E}_{b}
    -\eta_{AB}\accentset{(1)}{\Omega}^{Aa}\wedge \accentset{(0)}{\tau}^{B}\, ,
  \\
  \accentset{(1)}{R}^{Aa}(G)
  &
    = d\accentset{(1)}{\Omega}^{Aa}
    -\epsilon^{ab}\accentset{(0)}{\Omega}\wedge\accentset{(1)}{\Omega}^{A}_{\
    b}
    +\epsilon^{AB}\accentset{(0)}{\Omega}\wedge\accentset{(1)}{\Omega}^{\ a}_{B}\, .
\end{align}
\end{subequations}

Substituting the above expansions into the first-order Einstein-Hilbert
Lagrangian density 4-form
\begin{align}
  B
  &
    = \epsilon_{\hat{A}\hat{B}\hat{C}\hat{D}}R^{\hat{A}\hat{B}}
    \wedge E^{\hat{C}}\wedge E^{\hat{D}}
    \nonumber \\[7pt]
  &
  =2\epsilon_{cd}R(M)\wedge E^{c}\wedge E^{d}
  +\epsilon_{CD}R(J)\wedge \tau^{C}\wedge \tau^{D}
     -2\epsilon_{AB}\epsilon_{ab}R^{Aa}(G)\wedge \tau^{B}\wedge E^{b}\, .
\end{align}
we find that the  lowest-order term in the expansion
\begin{equation}
  \accentset{(0)}{B}
  = 2\epsilon_{AB}\accentset{(0)}{R}(J)\wedge \accentset{(0)}{\tau}^{A}
  \wedge\accentset{(0)}{\tau}^{B}\, .
\end{equation}
is, as expected, the action of 4D string Galilei gravity whose underlying
algebra is the string Galilei algebra discussed in the introduction.

The next order term leading to an invariant action is given by
\begin{align}
  \accentset{(2)}{B}
     & =  2\epsilon_{ab}\accentset{(0)}{R}(M)\wedge \accentset{(1)}{E}^{a}\wedge
    \accentset{(1)}{E}^{b}
    +2\epsilon_{AB}\left[\accentset{(2)}{R}(J) \wedge
    \accentset{(0)}{\tau}^{A}
    +2 \accentset{(0)}{R}(J) \wedge \accentset{(2)}{\tau}^{A}\right]
    \wedge\accentset{(0)}{\tau}^{B} \nonumber\\[2pt]
    &\,\,\,\,\,\, -2\epsilon_{AB}\epsilon_{ab}\accentset{(1)}{R}^{Aa}(G)\wedge\accentset{(0)}{\tau}^{B}
    \wedge \accentset{(1)}{E}^{b}\, .
\end{align}

Renaming the fields that appear at second order as
\begin{multicols}{3}
\begin{subequations}
\setlength{\abovedisplayskip}{-15pt}
\allowdisplaybreaks
\begin{align}
\accentset{(2)}{\tau}^{A}   & =  m^{A}\, , \\
\accentset{(2)}{\Sigma}   & =  n\, , \\
\accentset{(2)}{\Omega}   & =  s\, ,
\end{align}
\end{subequations}
\end{multicols}
\noindent
and using for the lowest-order components the  label of the original fields
we obtain the following expression for $\accentset{(2)}B$:\,\footnote{As in the first example, this action has extra accidental symmetries \cite{Harmark:2018cdl} leading to the enlarged algebra given in \cite{inpreparation}.}
\begin{align}
  \accentset{(2)}{B}
  &
    = 2\epsilon_{ab}R(M)\wedge E^{a}\wedge E^{b}+2\epsilon_{AB}R(s)
    \wedge \tau^{A} \wedge\tau^{B}
    \nonumber \\[6pt]
  &
    \,\,\,\,\,\,
    +4\epsilon_{AB} R(J) \wedge m^{A} \wedge\tau^{B}
    -2\epsilon_{AB}\epsilon_{ab}R^{Aa}(G)\wedge\tau^{B}\wedge E^{b}\, ,
\end{align}
which is precisely the Lagrangian density corresponding to the action of
extended string Newton-Cartan theory constructed in \cite{Bergshoeff:2018vfn}.

.

\section{Discussion}
\label{sec-discussion}

In this work we have shown that the procedure of Lie algebra expansions
provides a systematic method for constructing Lie algebras and actions for
non-relativistic gravity. All examples we have given are characterized by 4
numbers $(D,p;N_{0},N_{1})$ where $(D,p)$ refers to the fact that the
corresponding non-relativistic gravity theory naturally couples to a $p$-brane
moving in $D$ target space dimensions and $(N_{0},N_{1})$, with
$N_{0}=N_{1}\pm 1$, defines the truncation of the Lie algebra expansion.

We find that, in all cases, the lowest order term in the expansion,
$(D,p;N_{0},N_{1}) = (D,p;0,1)$, leads to the $(D,p)$ Galilei algebra,
\textit{i.e.},~a $D$-dimensional Galilei algebra with $p+1$ longitudinal
translation generators, with a corresponding action for Galilei gravity. These
Galilei gravity theories have been studied in \cite{Bergshoeff:2017btm}. They
are a bit special in the sense that not all gauge fields corresponding to the
generators of the Galilei algebra occur in the action\footnote{In the 3D case,
  this corresponds to the fact that, unlike the Poincar\'e algebra, the
  Galilei algebra has no invariant non-degenerate bilinear form.}.  The
next-order algebra, $(D,p;N_{0},N_{1}) = (D,p;2,1)$, does not necessarily
allow for an invariant action for arbitrary values of $D$ and $p$. We have
given three examples (Examples~1, 2 and 5 in section~5) where an invariant
action did exist, and one example (Example~4 in section~5) where, for the
purpose of constructing an invariant action, one needs to go to one order
higher, \textit{i.e.},~one needs to consider the truncation
$(N_{0},N_{1})= (2,3)$. We have furthermore studied an example (Example~3 in section~5) with a
truncation at order $(N_{0},N_{1})= (4,3)$.

It is a general feature of the Lie algebra expansion that the lowest-order
$(D,p)$ Galilei algebra can be obtained as a In\"on\"u-Wigner contraction of
the Poincar\'e algebra. Similarly, the action for the $(D,p)$ Galilei gravity
can be obtained as a $c\rightarrow \infty$ limit of the first-order Einstein-Hilbert action
\cite{Bergshoeff:2017btm}.\footnote{The expansion of the Vierbein  in terms of our  parameter $\lambda$ is the same as the usual expansion in terms of  the parameter $c^{-1}$, see,
  \textit{e.g.},~\cite{VandenBleeken:2017rij}, provided we multiply this
  Vierbein, before expanding,  with a factor $c$.}  It is intriguing that the
actions in two of the examples we have given, \textit{i.e.}, 3D extended
Bargmann gravity and 4D extended string Newton-Cartan gravity, can also be
obtained as the non-relativistic limit of the first order Einstein-Hilbert
Lagrangian even though they correspond to a next-order term in the Lie algebra
expansion.  However, in order to define a regular limit, one needs to add an
extra term to the Einstein-Hilbert action that contains two vectors in the
case of 3D extended Bargmann gravity \cite{Bergshoeff:2016lwr} and a vector
plus 2-form gauge field in the case of 4D extended string Newton-Cartan
gravity \cite{Bergshoeff:2018vfn}. These two cases can be generalized to a
$p$-brane moving in $D=p+3$ dimensions in which case one has to add an extra
term to the Einstein-Hilbert term involving a vector and a $(p+1)$-form gauge
field \cite{Bergshoeff:2018vfn}. The corresponding extended $p$-brane
Newton-Cartan algebras that can be used in the Lie algebra expansion have been
given in \cite{Bergshoeff:2018vfn}.\footnote{The common feature of all these
  $p$-branes is that they have a two-dimensional transverse space allowing
  anyonic extended objects like anyonic particles in three spacetime
  dimensions.
  }

The purpose of the extra term is to cancel a divergence that arises from the
lowest-order term in the Lie-algebra expansion which for this class of models
is proportional to
 \begin{equation}
 \epsilon_{abC_{1}\cdots C_{P+1}} R^{ab}(J)\wedge \tau^{C_{1}}
\wedge \cdots \wedge \tau^{C_{P+1}}\, .
\end{equation}

What makes this term special is that, for $D=p+3$, both $[ab]$ and
$[C_{1}\dots C_{P+1}]$ are singlets under the longitudinal and transverse
Lorentz rotations. This makes it possible to cancel the divergence that
follows from this term by adding the following term to the Einstein-Hilbert
action \cite{Bergshoeff:2018vfn}:
 \begin{equation}
   \epsilon^{\mu_{1}\cdots \mu_{p+3}}
   \big(\partial_{\mu_{1}}{\hat{A}}_{\mu_{2}}\big) {\hat{B}}_{\mu_3\cdots \mu_{p+3}}\,.
 \end{equation}

It remains to be seen whether and how the third and fourth example can also be obtained
as a special limit of General Relativity. The algebra underlying the fourth example
is precisely the algebra occurring in \cite{Hansen:2018ofj}\,\footnote{In the second reference of  \cite{Hansen:2018ofj} it is noted that the same algebra can be obtained following  \cite{Khasanov:2011jr} in which  Lie algebra expansions from a somewhat different perspective are discussed.}, suggesting that,
perhaps, the action we constructed and the one of \cite{Hansen:2018ofj} are
the same.\,\footnote{Both the action of  \cite{Hansen:2018ofj} and the action we constructed can be formulated in any dimension $D>3$. In our case, it requires that we take $n=D-1$.} As it is the case here, in \cite{Hansen:2018ofj} an expansion of
both the diffeomorphisms and the Lorentz gauge parameters is used, so it would
be interesting to work out the connection (if any) with the expansions
considered here. Note that we are working here with a first-order
formulation, whereas the action of \cite{Hansen:2018ofj} is presented in a
second-order formulation. The fact that in such a second-order formulation the
spin-connection fields are dependent makes the connection to the underlying
algebra less straightforward. Another difference is that the geometry of
\cite{Hansen:2018ofj} has twistless torsion whereas a priori we do not have such a restriction.
However, it is known from
examples 1 and 5 that going to a second-order formulation {\it after} the expansion
one cannot always solve for all the spin connection fields.  The spin
connection fields that cannot be solved for act as Lagrange multipliers
imposing geometric constraints. For instance, in the case of 4D Galilei
gravity, the independent spin connection imposes the constraint that the
geometry has twistless torsion (see eq.~\eqref{eq: RHab is zero} above or
eq.~(4.28) of \cite{Bergshoeff:2017btm}). Perhaps, a similar thing happens in
the fourth example in which case the corresponding spin connection could play
the same role as the Lagrange multiplier that is added by hand in
\cite{Hansen:2018ofj}. This remains to be investigated.

It is not clear how to couple the non-relativistic gravity theories
constructed in this work to a particle (examples 1-4) or a string (example 5)
preserving all the symmetries. The Galilei algebra lacks the central charge
symmetry that is needed to construct a massive particle action invariant under
Galilean boosts\,\footnote{To obtain a massive particle action invariant under Galilean boosts one needs to couple the particle to the central charge gauge field via a Wess-Zumino term.} while the other algebras have extra symmetries whose physical
interpretation is not clear and which make them non-trivial to realize on
matter.\footnote{For some efforts in this direction in the case of the 3D
  extended Bargmann algebra, see \cite{Alvarez:2007fw}. For a recent proposal in the context of example 4,
  see the second reference in \cite{Hansen:2018ofj}.}  In the absence of
such matter couplings we prefer to reserve the name Newton-Cartan gravity for
the Bargmann algebra only and to call the non-relativistic gravity theories
that are based on the larger algebras that occur in the Lie algebra expansion
`extended Newton-Cartan gravity'.  Sticking to this definition, it remains to
be seen whether an action for Newton-Cartan gravity can be constructed.
The same issues arise in the string case where the 4D string NC background that
couples to the Polyakov string has less symmetries than the 4D {\it extended} string NC gravity
that occurs in the last example.

An attractive feature of the Lie algebra expansion method is that it is a
quite general construction procedure and can be applied to a variety of new
situations.  One natural case to consider is actions for (conformal extensions
of) Carroll gravity corresponding to taking ultra-relativistic limits. We
expect that in this case the lowest order term in the expansion of the
Einstein-Hilbert action will correspond to the Carroll gravity action of
\EB{\cite{Henneaux:1979vn,Bergshoeff:2017btm}}.\,\footnote{The actions of \cite{Henneaux:1979vn} and \cite{Bergshoeff:2017btm} look rather different using different variables. Since they are both based upon the same Carroll symmetries, we expect them to coincide up to redefinitions.}  One can also apply the Lie algebra expansion method to extensions of the  Poincaré algebra such as the 3D higher-spin algebras considered in \cite{Bergshoeff:2016soe}. The same method can also be used to construct new
examples of supersymmetric extensions of non-relativistic gravity theories
beyond the known example of the supersymmetric extension of 3D extended
Bargmann gravity \cite{Bergshoeff:2016lwr,Izquierdo}.
We hope to come back soon to these interesting issues.

\section*{Acknowledgments}

EB~wishes to thank Joaquim Gomis, Kevin Grosvenor, Johannes Lahnsteiner, Jan Rosseel, Ceyda \c{S}im\c{s}ek and Ziqi Yan
for useful discussions. JMI wishes to thank J.A.~de Azc\'arraga and
D. G\'utiez for useful conversations. This work and LR have been supported in
part by the MINECO/FEDER, UE grant FPA2015-66793-P and by the Spanish Research
Agency (Agencia Estatal de Investigaci\'on) through the grant IFT Centro de
Excelencia Severo Ochoa SEV-2016-0597. JMI has been supported by the grants
MTM2014-57129-C2-1-P from the MINECO (Spain), VA137G18 Spanish Junta de
Castilla y Le\'on and FEDER BU229P18. TO wishes to thank M.M.~Fern\'andez for
her permanent support.  EB wishes to thank the Instituto de F\'{\i}sica
Te\'orica UAM/CSIC of Madrid for hospitality.

\appendix

\section{Deriving Invariance Conditions of the Action of the
  $(D,p;N_{0},N_{1})$ Algebra}
\label{app-invarianceconditions}

In this appendix we give a derivation of the invariance conditions
\eqref{invcond} that have to be satisfied for the action of a
$(D,p;N_{0},N_{1})$ non-relativistic gravity theory to be Lorentz
invariant.\footnote{We do not consider the $P$-transformations
generated by the $P_{\hat A}$ generators of the Poincar\'e algebra since, as
  explained in the main text, they are represented by the general coordinate
  transformations that are manifestly realized.}  WE remind the reader that
$(D,p)$ refers to the fact that the corresponding non-relativistic gravity
theory naturally couples to a $p$-brane moving in $D$ target space dimensions
and $(N_{0},N_{1})$ with $N_{0}=N_{1}\pm 1$ defines the truncation of the Lie
algebra expansion.

Since the starting action is invariant under Lorentz transformations, in an
infinite expansion each term order by order will be invariant under all the
generators that arise in the infinite expansion of the Lorentz
ones.\footnote{These include Lorentz transformations in the longitudinal
  directions, spatial rotations in the transverse directions plus Galilean
  boosts.}  However, if we perform a truncation to to the orders
$\mathfrak{g}(N_{0},N_{1})$, some of the terms at a given order may not
appear, thereby spoiling the invariance of the action. We will now discuss the
conditions under which this does not happen and the invariance is not broken.

A term in the expansion at order $\accentset{(n)}{B}$ does not appear if the
gap between the minimum order of the expansion, that we call $n_{min}$, and
$\accentset{(n)}{B}$ cannot be filled by a single field. In order to make this
argument more clear let us consider an example. If we have a Lagrangian
density $B$ given by the product of two fields, $a$ and $b$, $B=ab$ then
considering the full generic expansion
($a=\sum_{k=0}^{\infty}\ \accentset{(k)}{a}$ and
$b=\sum_{k=0}^{\infty}\ \accentset{(k)}{b}$) at order $n$, we get the
following expression
\begin{equation}
  \accentset{(n)}{B}=\sum_{k=0}^{n}
  \quad
  \accentset{(n-k)}{a}\, \,\,\,\,\, \accentset{(k)}{b}\, .
  \label{eq:exp1}
\end{equation}

After the truncation, the expansion of both fields $a$ and $b$ contains
terms to a maximum order, say $n_{a}$ and $n_{b}$ respectively, then the term
above should be written as
\begin{align}
  &\accentset{(n)}{B}=\sum_{k=0}^{\min\{n_{b}, n-n_{a}\}}\ \accentset{(n-k)}{a}
    \quad
    \accentset{(k)}{b}\, .
    \label{eq:exp2}
\end{align}
Thus, if
\begin{equation}
\min\{n_{b}, n-n_{a}\}<n,
\end{equation}
then \autoref{eq:exp1} will not contain all the terms appearing in
\autoref{eq:exp2}. This argument can be straightforwardly generalized to a
Lagrangian term containing more fields, as in our case.

In our case, the condition discussed above can be translated into the
following
\begin{equation}\label{eq:invariancecondition0}
N_{0}<n-n_{min}\,,\hskip 1.5truecm
N_{1}-1<n-n_{min},
\end{equation}
where the $-1$ in the second takes into account the fact that the minimum
order for $V_{1}$ is 1. These two inequalities can be written as one
inequality as follows:\footnote{With some caveat, see below.}
\begin{equation}
\max\{N_{0},N_{1}-1\}<n-n_{min}\,.
\end{equation}

Therefore,  in order to preserve the invariance of the action, we must require
\begin{equation}
  \label{eq:invariancecondition}
n\leqslant \min\{N_{0},N_{1}-1\}+n_{min}\, .
\end{equation}
However, this condition applies to a general term containing both fields from
$V_{0}$ and $V_{1}$. As we are going to discuss there could be certain values
of $D$ and $p$ for which a term in the Lagrangian contains only fields of one
of the subspaces. For those cases eq.~(\ref{eq:invariancecondition}) is too
strong.  Instead, only one of the conditions given in
eqs.~(\ref{eq:invariancecondition0}) should be satisfied.

To derive the invariance conditions, it is convenient to decompose the
expanded Lagrangian Eq.~(\ref{eq:D-dimexpandedLagrangian}) into three terms:
\begin{subequations}
\begin{align}
  B_{1}
  & =
    \binom{D-2}{p-1}R^{A_{0} A_{1}}(J)\wedge \tau^{A_{2}}\wedge \ldots\wedge \tau^{A_{p}}
    \wedge E^{a_{p+1}}\wedge \ldots\wedge E^{a_{D-1}}\, ,
  \\[4pt]
  B_{2}
  & =
    \binom{D-2}{p+1}R^{a_{p+1} a_{p+2}}(J)\wedge \tau^{A_{0}}\wedge \ldots\wedge
    \tau^{A_{p}}\wedge E^{a_{p+3}}\wedge \ldots\wedge E^{a_{D-1}}\, ,
  \\[4pt]
  B_{3}
  & =
    +2(-1)^{p}\binom{D-2}{p}R^{A_{0} a_{p+1}}(G)\wedge \tau^{A_{1}}\wedge
    \ldots\wedge \tau^{A_{p}}\wedge E^{a_{p+2}}\wedge \ldots\wedge E^{a_{D-1}}\, .
\end{align}
\end{subequations}
See Table~\ref{table:table1} for more information. Note that although
formally $B_{2}$ is defined for $p=-1$, its expansion is not well defined since
for $p=-1$ $R^{a_{p+1}a_{p+2}}$ would have been in $V_{0}$ but there is no
$V_{0}$. For this reason we do not consider the case $p=-1$.

\vskip .3truecm

\begin{table}[h!]
\renewcommand{\arraystretch}{1.5}
\begin{center}
\resizebox{\textwidth}{!}{
\begin{tabular}{|c|c|c|c|c|c|}
\hline
\bf Term &\bf Min Order& \bf Max Order&\bf Existence& \bf Only $\mathbf{V_{0}}$ Terms&\bf Only $\mathbf{V_{1}}$ Terms\\
\hline
$B_{1}$&$D-p$&$(D-p-1)N_{1}+pN_{0}$&$p\geqslant 1$&$D=p+1$&-\\
$B_{2}$&$D-p-3$&$(D-p-3)N_{1}+pN_{0}$&$p\leqslant D-3$&$D=p+3$&-\\
$B_{3}$&$D-p-1$&$(D-p-1)N_{1}+pN_{0}$&$0\leqslant p \leqslant D-2$&-&$p=2$\\
\hline
\end{tabular}
}
\end{center}
\caption[]{This table indicates the minimum and maximum order of each term,
  the existence conditions, that are also depicted in
  Fig.~\ref{fig:existenceconditions}, and the conditions under which each term
  contains only fields coming from one of the subspaces.
}\label{table:table1}
\end{table}
\begin{figure}
  \begin{center}
    \includegraphics[scale=1.5]{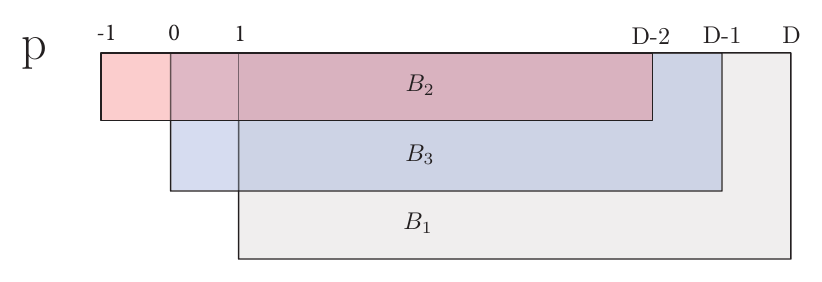}
    \caption{Existence conditions of the terms $B_{1},\ B_{2}$ and $B_{3}$ in
      the action that could be read from Table~\ref{table:table1}.}
      \label{fig:existenceconditions}
  \end{center}
\end{figure}

In order to find the invariance conditions we must evaluate
eq.~(\ref{eq:invariancecondition}) for each term in the different existence
regions, taking into account the special values of $D$ and $p$ for which the
terms we are considering contain only fields from one of the subspaces $V_{0}$
or $V_{1}$. For instance, if we take $p=0$ we see from
eqs.~(\ref{fig:existenceconditions}) that only the terms $B_{2}$ and $B_{3}$
exist and that $B_{3}$ has only contributions from $V_{1}$. Thus, the
invariance conditions read
\begin{subequations}
\begin{align}
n\leqslant N_{1}+D-1.
\end{align}

For $B_{2}$ instead we should distinguish  the following two cases:%
\begin{align}
\left\{
\begin{array}{lcl}
D=3 &&n \leqslant \min\{N_{0},N_{1}+1\}\,,\\ [0.2cm]
D\neq 3 &&n\leqslant \min\{N_{0},N_{1}-1\}+D-3\,.
\end{array}\right.
\end{align}

Putting together these two conditions we deduce that for $p=0$ the term
$\accentset{(n)}{B}$ is invariant provided that
\begin{align}
\left\{
\begin{array}{lcl}
D=3 &&n \leqslant \min\{N_{0},N_{1}+1\}\,,\\ [0.2cm]
D\neq 3 &&n\leqslant \min\{N_{0},N_{1}-1\}+D-3\,.
\end{array}\right.
\end{align}
\end{subequations}

Studying in a similar way the different invariance conditions for different
values of $p$, we deduce the following conditions:
\begin{subequations}
\begin{align}
&p=0&&\left\{
\begin{array}{lcl}
D=3 &&n \leqslant \min\{N_{0},N_{1}+1\}\,,\\ [0.2cm]
D\neq 3 &&n\leqslant \min\{N_{0},N_{1}-1\}+D-3\,,
\end{array}\right.\\
&\nonumber\\
&1 \leqslant p\leqslant D-3&&\left\{
\begin{array}{lcl}
D=p+3 &&n \leqslant \min\{N_{0},\min\{N_{0},N_{1}-1\}+2\}\,,\\ [0.2cm]
D\neq p+3 &&n\leqslant \min\{N_{0},N_{1}-1\}+D-p-3\,,\\
\end{array}\right.\\
&\nonumber\\
&p=D-2&&n\leqslant \min\{N_{0},N_{1}-1\}+1\,,\\
&\nonumber\\
&p=D-1&&n\leqslant N_{0}\,.
\end{align}
\end{subequations}

\noindent
We note that the case $p=-1$ is not realized. For the two allowed cases
$N_{1}\pm 1=N_{0}$ the conditions above can be simplified, using the identity
\begin{equation}
\min\{N_{0},N_{1}-1\}=\min\{N_{1}\pm 1,N_{1}-1\}=N_{1}-1\,,
\end{equation} as follows:
\begin{subequations}
\begin{align}
&p=0&&\left\{
\begin{array}{lcl}
D=3 &&n \leqslant N_{0}\\ [0.2cm]
D\neq 3 &&n\leqslant N_{1}+D-4
\end{array}\right.\\
&\nonumber\\
&1 \leqslant p\leqslant D-3&&\left\{
\begin{array}{lcl}
D=p+3 &&n \leqslant N_{0}\\ [0.2cm]
D\neq p+3 &&n\leqslant N_{1}+D-p-4\\
\end{array}\right.\\
&\nonumber\\
&p=D-2&&n\leqslant N_{1}\\
&\nonumber\\
&p=D-1&&n\leqslant N_{0}
\end{align}\label{eq:Lorentzinvappendix}
\end{subequations}

As an example, we give in Table~\ref{table:table3} the solution of the
invariance conditions for $D=5$ and different values of $p$ corresponding to
the truncated algebras $\mathfrak{g}(2,1)$ and $\mathfrak{g}(2,3)$:

\begin{table}[!ht]
\renewcommand{\arraystretch}{1.5}
\begin{center}
\begin{tabular}{|c|c|c|c|c|c|}
\hline
\bf Algebra&\bf $\mathbf{p=0}$&\bf $\mathbf{p=1}$&\bf $\mathbf{p=2}$&\bf $\mathbf{p=3}$&\bf $\mathbf{p=4}$ \\
\hline
$\mathfrak{g}(2,1)$&$n\leqslant 2$&$n\leqslant 1$&$n\leqslant 2$&$n\leqslant 1$&$n\leqslant 2$\\
$\mathfrak{g}(2,3)$&$n\leqslant 4$&$n\leqslant 3$&$n\leqslant 2$&$n\leqslant 3$&$n\leqslant 2$\\
\hline
\end{tabular}
\end{center}
\caption[]{\small Invariance conditions for the expanded terms
  $\accentset{(n)}{B}$ under the algebras $\mathfrak{g}(2,1)$ and
  $\mathfrak{g}(2,3)$.}\label{table:table3}
\end{table}

\section{Transformation rules and curvatures of Example 4}
\label{appendix}

For the convenience of the reader, we give in this appendix the definition of
the curvatures and the transformation rules of the gauge fields plus
curvatures that occur in Example 4, i.e.~the case
$(D,p;N_{0},N_{1})= (4,0;2,3)$, discussed in the main text.

The transformation rules for the fields corresponding to the generators of the
$\mathfrak{g}(2,3)$ algebra are given by
\begin{subequations}\label{transfgf}
\begin{align}
  \delta \accentset{(1)}{E}^{a}
  &
    =
    d\accentset{(1)}{\epsilon}^{a}
    -\accentset{(1)}{\lambda}^{a}\accentset{(0)}{\tau}
    +\accentset{(0)}{\lambda}\accentset{(1)}{\Omega}^{a}
    -\accentset{(0)}{\lambda}^{ab}\accentset{(1)}{E}_{b}
    +\accentset{(1)}{\epsilon}_{b}\accentset{(0)}{\Omega}^{ab}\, ,
  \\
  \delta \accentset{(3)}{E}^{a}
  &
    = d\accentset{(3)}{\epsilon}^{a}
    -\accentset{(3)}{\lambda}^{a}\accentset{(0)}{\tau}
    +\accentset{(0)}{\lambda}\accentset{(3)}{\Omega}^{a}
    -\accentset{(1)}{\lambda}^{a}\accentset{(2)}{\tau}
    +\accentset{(2)}{\lambda}\accentset{(1)}{\Omega}^{a}
    -\accentset{(0)}{\lambda}^{ab}\accentset{(3)}{E}_{b}
    +\accentset{(3)}{\epsilon}_{b}\accentset{(0)}{\Omega}^{ab}
    -\accentset{(2)}{\lambda}^{ab}\accentset{(1)}{E}_{b}
    +\accentset{(1)}{\epsilon}_{b}\accentset{(2)}{\Omega}^{ab}\, ,
  \\
  \delta \accentset{(1)}{\Omega}^{a}
  &
    = d\accentset{(1)}{\lambda}^{a}
    -\accentset{(0)}{\lambda}^{ab}\accentset{(0)}{\Omega}_{b}
    +\accentset{(1)}{\lambda}_{b}\accentset{(0)}{\Omega}^{ab}\, ,
  \\
  \delta \accentset{(3)}{\Omega}^{a}
  &
    = d\accentset{(1)}{\lambda}^{a}
    -\accentset{(0)}{\lambda}^{ab}\accentset{(0)}{\Omega}_{b}
    +\accentset{(1)}{\lambda}_{b}\accentset{(0)}{\Omega}^{ab}
    -\accentset{(2)}{\lambda}^{ab}\accentset{(1)}{\Omega}_{b}
    +\accentset{(1)}{\lambda}_{b}\accentset{(2)}{\Omega}^{ab}\, ,
  \\
  \delta \accentset{(0)}{\Omega}^{ab}
  &
    = d\accentset{(0)}{\lambda}^{ab}
    +\accentset{(0)}{\lambda}^{ad}\accentset{(0)}{\Omega}^{b}_{\phantom{b}d}
    -\accentset{(0)}{\lambda}^{bd}\accentset{(0)}{\Omega}^{a}_{\phantom{b}d}\,  ,
  \\
  \delta \accentset{(2)}{\Omega}^{ab}
  &
    = d\accentset{(2)}{\lambda}^{ab}
    +\accentset{(0)}{\lambda}^{ad}\accentset{(2)}{\Omega}^{b}_{\phantom{b}d}
    -\accentset{(0)}{\lambda}^{bd}\accentset{(2)}{\Omega}^{a}_{\phantom{b}d}
    +\accentset{(2)}{\lambda}^{ad}\accentset{(0)}{\Omega}^{b}_{\phantom{b}d}
    -\accentset{(2)}{\lambda}^{bd}\accentset{(0)}{\Omega}^{a}_{\phantom{b}d}
    -2\accentset{(1)}{\lambda}^{[a}\accentset{(1)}{\Omega}^{b]}\, ,
  \\
  \delta \accentset{(0)}{\tau}
  &
    = d\accentset{(0)}{\lambda}\, ,
  \\
  \delta \accentset{(2)}{\tau}
  &
    = d\accentset{(2)}{\lambda}
    -\accentset{(1)}{\lambda}^{a}\accentset{(1)}{E}_{a}
    +\accentset{(1)}{\epsilon}^{a}\accentset{(1)}{\Omega}_{a}\, .
\end{align}
\end{subequations}
The curvatures that transform covariantly under these gauge transformations are given by
\begin{subequations}\label{curvatures2}
\begin{align}
  \accentset{(0)}{R}^{ab}(J)
  & =
    d\accentset{(0)}\Omega^{ab}-\accentset{(0)}\Omega^{ac}\wedge
    \accentset{(0)}\Omega^{b}_{\phantom{b}c}\, ,
  \\
   \accentset{(2)}R^{ab}(J)
  & =
    d\accentset{(2)}\Omega^{ab}-\accentset{(0)}\Omega^{ac}\wedge
    \accentset{(2)}\Omega^{b}_{\phantom{b}c} \accentset{(2)}\Omega^{ac}\wedge
    \accentset{(2)}\Omega^{b}_{\phantom{b}c} -\accentset{(1)}\Omega^{a}\wedge\accentset{(1)}\Omega^{b}\, ,
  \\
  \accentset{(1)}R^{a}(G)
  & =
    d\accentset{(1)}\Omega^{a}+
    +\accentset{(0)}\Omega^{ab}\wedge \accentset{(1)}\Omega^{a}\, ,
  \\
  \accentset{(3)}R^{a}(G)
  & =
    d\accentset{(3)}\Omega^{a}+
    +\accentset{(0)}\Omega^{ab}\wedge \accentset{(3)}\Omega^{a}
    +\accentset{(2)}\Omega^{ab}\wedge \accentset{(1)}\Omega^{a}\, ,
  \\
  \accentset{(1)}R^{a}(P)
  & =
    d\accentset{(1)}E^{a}+\accentset{(0)}\Omega^{ab}\wedge \accentset{(1)}{E}_{b}
    -\accentset{(1)}\Omega^{a}\wedge \accentset{(0)}{\tau}\, ,
  \\
  \accentset{(3)}R^{a}(P)
  & =
    d\accentset{(3)}E^{a}+\accentset{(0)}\Omega^{ab}\wedge \accentset{(3)}E_{b}
    +\accentset{(2)}\Omega^{ab}\wedge \accentset{(1)}{E}_{b}
    -\accentset{(3)}\Omega^{a}\wedge \accentset{(0)}{\tau}
    -\accentset{(1)}\Omega^{a}\wedge \accentset{(2)}{\tau}\, ,
  \\
  \accentset{(0)}{R}(H)
  & =
    d\accentset{(0)}{\tau}\, ,
 \\
  \accentset{(2)}R(H)
  & =
    d\accentset{(2)}{\tau}+\accentset{(1)}\Omega^{a}\wedge \accentset{(1)}{E}_{a}\,   .
\end{align}
\end{subequations}
These curvatures  transform as follows under the gauge transformations \eqref{transfgf} of the gauge fields:
\begin{subequations}
\begin{align}
  \delta \accentset{(1)}{R}^{a}(P)
  &
    = -\accentset{(1)}{\lambda}^{a}\accentset{(0)}{R}(H)
    +\accentset{(0)}{\lambda}\accentset{(1)}{R}^{a}(G)
    -\accentset{(0)}{\lambda}^{ab}\accentset{(1)}{R}_{b}(P)
    +\accentset{(1)}{\epsilon}_{b}\accentset{(0)}{R}^{ab}(J)\, ,
  \\
  \delta \accentset{(3)}{R}^{a}(P)
  &
    = -\accentset{(3)}{\lambda}^{a}\accentset{(0)}{R}(H)
    +\accentset{(0)}{\lambda}\accentset{(3)}{R}^{a}(G)
    -\accentset{(1)}{\lambda}^{a}\accentset{(2)}{R}(H)
    +\accentset{(2)}{\lambda}\accentset{(1)}{R}^{a}(G)
    -\accentset{(0)}{\lambda}^{ab}\accentset{(3)}{R}_{b}(P)
    +\accentset{(3)}{\epsilon}_{b}\accentset{(0)}{R}^{ab}(J)
    \nonumber \\
  &
    -\accentset{(2)}{\lambda}^{ab}\accentset{(1)}{R}_{b}(P)
    +\accentset{(1)}{\epsilon}_{b}\accentset{(2)}{R}^{ab}(J)\, ,
  \\
  \delta \accentset{(1)}{R}^{a}(G)
  &
    = -\accentset{(0)}{\lambda}^{ab}\accentset{(1)}{R}_{b}(G)
    +\accentset{(1)}{\lambda}_{b}\accentset{(0)}{R}^{ab}(J)\, ,
  \\
  \delta \accentset{(3)}{R}^{a}(G)
  &
    = -\accentset{(0)}{\lambda}^{ab}\accentset{(3)}{R}_{b}(G)
    +\accentset{(3)}{\lambda}_{b}\accentset{(0)}{R}^{ab}(J)
    -\accentset{(2)}{\lambda}^{ab}\accentset{(1)}{R}_{b}(G)
    +\accentset{(1)}{\lambda}_{b}\accentset{(2)}{R}^{ab}(J)\, ,
  \\
  \delta \accentset{(0)}{R}^{ab}(J)
  &
    = \accentset{(0)}{\lambda}^{ad}\accentset{(0)}{R}^{b}_{\phantom{b}d}(J)
    -\accentset{(0)}{\lambda}^{bd}\accentset{(0)}{R}^{a}_{\phantom{b}d}(J)\, ,
  \\
  \delta \accentset{(2)}{R}^{ab}(J)
  &
    = \accentset{(0)}{\lambda}^{ad}\accentset{(2)}{R}^{b}_{\phantom{b}d}(J)
    -\accentset{(0)}{\lambda}^{bd}\accentset{(2)}{R}^{a}_{\phantom{b}d}(J)
    +\accentset{(2)}{\lambda}^{ad}\accentset{(0)}{R}^{b}_{\phantom{b}d}(J)
    -\accentset{(2)}{\lambda}^{bd}\accentset{(0)}{R}^{a}_{\phantom{b}d}(J)
    -2\accentset{(1)}{\lambda}^{[a}\accentset{(1)}{R}^{b]}(G)\, ,
  \\
  \delta \accentset{(0)}{R}(H)
  &
    = 0\, ,
  \\
  \delta \accentset{(2)}{R}(H)
  &
    = -\accentset{(1)}{\lambda}^{a}\accentset{(1)}{R}_{a}(P)+
    \accentset{(1)}{\epsilon}^{a}\accentset{(1)}{R}_{a}(G)\, .
\end{align}\label{eq:curvaturesB}
\end{subequations}

\end{document}